\documentclass[a4,12pt]{article}
\usepackage[sort&compress,square,comma]{natbib}
\usepackage[T1]{fontenc}
\usepackage{bm}
\usepackage{amsmath}
\usepackage{epsfig}
\newcommand{\gvec}[1]{\hbox{\boldmath$#1$\unboldmath}}
\newcommand{\nvec}[1]{|\gvec{#1}|}

\begin{document}
\begin{flushright}
LYCEN-2009-10 \\
CERN-PH-TH/2009-192\\
\end{flushright}
\begin{center}
{\bf \large A unified approach for nucleon knock-out, 
coherent and incoherent pion production in neutrino interactions with nuclei}\\[2ex]

M. Martini$^{1,2,3}$, M. Ericson$^{1,3}$, G. Chanfray$^1$ and J. Marteau$^1$\\[1ex] 
 
$^1$  Universit\'e de Lyon, Univ.  Lyon 1, 
 CNRS/IN2P3,\\ IPN Lyon, F-69622 Villeurbanne Cedex, France\\
$^2$ Universit\`a di Bari, I-70126 Bari, Italy\\
$^3$ Theory Group, Physics Department,\\ CERN, CH-1211 Geneva, Switzerland 

\begin{abstract}
We present a theory of neutrino interactions with nuclei aimed at the description 
of the partial cross-sections, namely quasi-elastic and multi-nucleon emission, 
coherent and incoherent single pion production. For this purpose, we use the theory of nuclear responses 
treated in the random phase approximation, which allows a unified description of these channels. 
It is particularly suited for the coherent pion production where collective effects are important 
whereas they are moderate in the other channels. 
We also study the evolution of the neutrino cross-sections with the mass number from carbon to calcium.
We compare our approach to the available neutrino experimental data on carbon. 
We put a particular emphasis on the multi-nucleon channel, which at present is not 
easily distinguishable from the quasi-elastic events. This component turns out to be quite 
relevant for the interpretation of experiments (K2K, MiniBooNE, SciBooNE). 
It can account in particular for the unexpected behavior of the quasi-elastic cross-section. 
\end{abstract}

\end{center}
%\vskip 0.5 cm 

\vskip 0.5 cm 

%\begin{document} 
%%%%%%%%%%%%%%%%%%%%%%%%%%%%%%%%%%%%%%%%%%%%%%%%%%%%%%%%%%%%%%%%%%%%%%%%%
\section{Introduction}
%%%%%%%%%%%%%%%%%%%%%%%%%%%%%%%%%%%%%%%%%%%%%%%%%%%%%%%%%%%%%%%%%%%%%%%%%
 Neutrino physics has undergone a spectacular development in the last decade, following 
the discovery of neutrino oscillations first revealed by the anomaly of 
atmospheric neutrinos \cite{Fukuda:1998mi}. A number of results on the interaction of neutrinos 
with matter are now available. Neutrino detectors do not usually consist of pure hydrogen but they 
involve complex nuclei for instance $^{12}$C, as in SciBar \cite{AguilarArevalo:2006se},
where the molecule C$_8$H$_8$ is involved, or in MiniBooNE \cite{AguilarArevalo:2008qa} 
which uses the mineral oil CH$_2$. Heavier nuclei are also under consideration for 
instance in the liquid argon chamber planned for T2K \cite{Itow:2001ee,Terri:2009zz}. 
A number of  results have been obtained,  for neutral or charged current 
(K2K, MiniBooNE, SciBooNE) on quasi elastic processes or coherent 
and incoherent single pion production 
\cite{Nakayama:2004dp,Hasegawa:2005td,Gran:2006jn,:2007ru,AguilarArevalo:2008xs,
:2008eaa,Hiraide:2008eu,AguilarArevalo:2008yp,AguilarArevalo:2009eb,
Katori:2009du,Kurimoto:2009yb,AlcarazAunion:2009ku}. 
The first question is then if our present 
understanding of neutrino interactions with matter can reproduce the available data. 
Many works 
\cite{Paschos:2005km,Singh:2006bm,Ahmad:2006cy,Rein:2006di,AlvarezRuso:2007tt,AlvarezRuso:2007it,Butkevich:2008ef,Praet:2008yn,SajjadAthar:2008hi,Amaro:2008hd,Leitner:2008wx,Berger:2008xs,Leitner:2009ph,Paschos:2009ag,Benhar:2009wi,Hernandez:2009vm,Butkevich:2009cp,Athar:2009rc,Nakamura:2009iq}
have been devoted to this problem, using various theoretical 
approaches \cite{Delorme:1985ps,Marteau:1999kt,Marteau:1999jp,Singh:1992dc,Singh:1993rg,AlvarezRuso:1997jr,Singh:1998ha,Sato:2003rq,Szczerbinska:2006wk,Paschos:2003qr,Meucci:2003cv,Meucci:2004ip,Nieves:2004wx,Amaro:2004bs,Caballero:2005sj,Amaro:2006tf,Martini:2007jw,Amaro:2006if,Ivanov:2008ng,Martinez:2005xe,Benhar:2005dj,Benhar:2006qv,Benhar:2006nr,Ankowski:2005wi,Leitner:2006ww,Leitner:2006sp,Hernandez:2007qq,Leitner:2008ue}. 
In this article we will explore such interactions using the 
theory of the nuclear response treated in the random phase approximation (RPA) 
in the quasi-elastic and Delta resonance region including also two and three nucleon knock-out. 
The formalism is the same as the one used by Marteau \cite{Marteau:1999kt} 
in his work on the $\nu-^{16}$O interaction.
The merit of this approach is that, although 
perfectible in several ways, it describes in a unique frame several final state channels.
This technique has been successful in a number of problems involving either 
weakly interacting probes such as (e,e') scattering or strongly interacting 
ones such as pion scattering or ($^{3}$He,T) charge exchange reaction \cite{Delorme:1989nh}. 
We give the cross-sections for pion production, coherent or incoherent, and nucleon knock-out, 
for neutral or charged currents. We restrict to single pion production 
ignoring two-pion production processes which, for real photons, lead to a sizable 
part of the photo-absorption cross-section at energies larger than the 
Delta resonance, above $\simeq500$ MeV. Our treatment should thus underestimate 
the cross-section  when multi-pion production starts to show up. 
Our work  ignores as well the meson exchange effects which play 
a non negligible role \cite{Alberico:1983zg,De Pace:2003xu}. 
We only take into account the exchange effect in the time component 
of the axial current, which is known to be important \cite{Kubodera:1978wr}.
For single pion production we assume that the dominant production  mechanism  
is via the Delta resonance, ignoring the other resonance excitations, which also 
limits the energy for the validity of our approach. 
Beyond quasi-elastic processes and single pion production via Delta 
excitation we also incorporate several nucleon 
knock-out through two-particle-two-hole ($2p-2h$) and $3p-3h$ excitations. 
These will play a crucial role in the comparison with data involving quasi-elastic events.

Among the aims of this work there is the exploration of the evolution of the neutrino-nucleus 
interaction as the mass number of the nucleus goes from the carbon region to 
the region of $^{40}$Ca. 
This investigation is motivated by the project of a liquid argon chamber in 
the T2K experiment which raises the question if one 
keeps control of the understanding of the interaction of neutrinos with matter by going 
to a medium-weight nucleus such as $^{40}$Ar. 
In order to single out the evolutions linked to the nuclear size we have 
chosen as element of comparison an isoscalar nucleus in the $^{40}$Ar region, 
namely  $^{40}$Ca.  For the coherent process which per nucleon fades away in heavy nuclei, 
the evolution is relatively rapid but should remain under control 
as our theory is particularly well adapted to this channel. 
The other exclusive channels, in particular the incoherent pion production, are 
sensitive to final state interaction not automatically included in our approach. 
This leaves some uncertainty in the evolution between the mass 12 and the mass 40 region for this channel. 

Our article is organized as follows: Section 2 introduces the formalism of the response 
functions treated in the random phase approximation (RPA). 
Section 3 discusses the various final state channels. 
In Sec. 4 we compare these predictions with the available data. In Sec. 5 we provide 
a summary and conclusion of the present work.

%%%%%%%%%%%%%%%%%%%%%%%%%%%%%%%%%%%%%%%%%%%%%%%%%%%%%%%%%%%%%%%%%%%%%%%%%%%%%%%%%  
\section{Formalism}     
%%%%%%%%%%%%%%%%%%%%%%%%%%%%%%%%%%%%%%%%%%%%%%%%%%%%%%%%%%%%%%%%%%%%%%%%%%%%%%%%%%
The double differential cross-section  for the reaction \mbox{$ \nu_l \, (\bar{\nu}_l) + A \longrightarrow l^- \, (l^+) + X $}
is given by
\begin{equation}
\label{m_eq_1}
\frac{\partial^2 \sigma}{\partial\Omega_{k'} \partial k'} = \frac{G_F^2 \cos^2\theta_C \gvec{k'}^2}{32\pi^2k_0{k'}_0} |T|^2 
\end{equation}
where $ G_F $ is the weak coupling constant, $ \theta_c $ the Cabbibo angle, 
$ k $ and $ k^\prime $ the initial and final lepton momenta, and $T$ the invariant amplitude given by the contraction of 
the leptonic $L$ and hadronic $W$ tensors. Their expressions are given in Appendix \ref{appendix A}.
In order to illustrate how the various response functions enter and to introduce the variables, we give below a simplified expression, 
which in particular ignores the lepton mass contribution 
and assumes zero $\Delta$ width. We stress however that in the actual calculations the full formulas of Appendix
\ref{appendix A}, 
which do not make these simplifications, have been used. The simplified double differential cross-section reads 
\begin{eqnarray} \label{SIGMANUENU}
\frac{\partial^2\sigma}{\partial\Omega \,\partial k^\prime} & = & \frac{G_F^2 \, 
\cos^2\theta_c \, (\gvec{k}^\prime)^2}{2 \, \pi^2} \, \cos^2\frac{\theta}{2} \, 
\left[ G_E^2 \, (\frac{q_\mu^2}{\gvec{q}^2})^2 \, R_\tau^{NN} \right. \nonumber \\ 
& + & G_A^2 \, \frac{( M_\Delta - M_N )^2}{2 \, \gvec{q}^2} \, R_{\sigma\tau (L)}^{N\Delta} +  G_A^2 \, \frac{( M_\Delta - M_N )^2}{\gvec{q}^2} R_{\sigma\tau (L)}^{\Delta\Delta} \nonumber \\ 
& + & \left( G_M^2 \, \frac{\omega^2}{\gvec{q}^2} + G_A^2 \right) \, 
 \left( - \frac{q_\mu^2}{\gvec{q}^2} + 2 \tan^2\frac{\theta}{2} \right) \,
\left( R_{\sigma\tau (T)}^{NN} + 2 R_{\sigma\tau (T)}^{N\Delta} 
+ R_{\sigma\tau (T)}^{\Delta\Delta} \right) \nonumber \\ 
& \pm & \left. 2 \, G_A \, G_M \, \frac{k + k^\prime}{M_N} \, 
\tan^2\frac{\theta}{2} \, 
\left( R_{\sigma\tau (T)}^{NN} + 2 R_{\sigma\tau (T)}^{N\Delta} 
+ R_{\sigma\tau (T)}^{\Delta\Delta} \right) \right] 
\end{eqnarray}
where 
$ q_\mu = k_\mu - k_\mu^\prime = ( \omega,\gvec{q} ) $ is the four momentum transferred to the 
nucleus, $ \theta $ the scattering angle, $ M_\Delta $ ($ M_N $) the Delta 
(nucleon) mass. 
The electric, magnetic and axial form 
factors are taken in the standard dipole parameterization with the following 
normalizations: $ G_E(0) = 1.0 $, $ G_M(0) = 4.71 $ and $ G_A(0) = 1.255 $. 
The corresponding cut-off parameters are $M_V=0.84~\textrm{GeV}/c^2$ for the electric and magnetic terms 
and $M_A=1.032~\textrm{GeV}/c^2$ for the axial one. The plus (minus) sign in Eq. (\ref{SIGMANUENU}) stands for the 
neutrino (antineutrino) case. 
A similar expression applies to the process~:
$ \nu_l \, (\bar{\nu}_l) + A 
\longrightarrow  \nu_l \, (\bar{\nu}_l)  + X $,
 which involves neutral currents. The various responses $R$ appearing in Eq.(\ref{SIGMANUENU}) are defined according to

\begin{equation} \label{eq:2}
R_\alpha^{PP^\prime} = \sum_n \, 
\langle n | \sum_{j=1}^A \, O_\alpha^P(j) \, 
e^{ i \, \gvec{q}.\gvec{x}_j } | 0 \rangle 
\langle n | \sum_{k=1}^A \, O_\alpha^{P^\prime}(k) \, 
e^{ i \, \gvec{q}.\gvec{x}_k } | 0 \rangle^* \, \delta (\omega - E_n + E_0 ).  
\end{equation}
The upper indices $(P,P')$ refer to the type of particle ($N$ or $\Delta$) at the vertices that couples to the external probe. 
The corresponding operators have the following forms:
$$ 
O_\alpha^N(j) = \tau_j^\pm, \,\,\, ( \gvec{\sigma}_j . \widehat{q} ) \, 
\tau_j^\pm, \,\,\, 
( \gvec{\sigma}_j \times \widehat{q} )^i
\, \tau_j^\pm, 
$$
for $ \alpha = \tau $,  $ \sigma\tau (L) $, $ \sigma\tau (T) $, and 
$$
O_\alpha^\Delta(j) = ( \gvec{S}_j . \widehat{q} ) \, T_j^\pm, \,\,\, 
%(( \gvec{S}_j \times \widehat{q} ) \times \widehat{q} ) \, T_j^\pm,
( \gvec{S}_j \times \widehat{q} )^i \, T_j^\pm, 
$$
for $ \alpha = \sigma\tau (L) $, $ \sigma\tau (T) $. We have thus defined the inclusive \textit{isospin} ($ R_\tau $), 
\textit{spin-isospin longitudinal} ($ R_{\sigma\tau (L)} $) and \textit{spin-isospin transverse} ($ R_{\sigma\tau (T)} $) 
nuclear response 
functions (the longitudinal and transverse character of these last two responses refers to the direction of the spin operator with respect to the 
direction of the transferred momentum). The operators $ S $ and $ T $ are the usual 1/2 to 3/2 transition operators in the spin and isospin space. We have assumed the existence of a scaling law between the nucleon and Delta magnetic and axial form factors 
\cite{chew/low}: 
$$
 G_M^* / G_M = G_A^* / G_A = f^* / f, 
$$ where $ f^* $ ($ f $) is the $ \pi \, 
N \, \Delta $ 
($ \pi \, N\, N $) coupling constant. For a matter of convenience, we have 
incorporated the scaling factor $ f^* / f  = 2.2 $ into the responses. 

\begin{figure}[ht]
\begin{center}
\includegraphics[width=12cm,height=7cm]{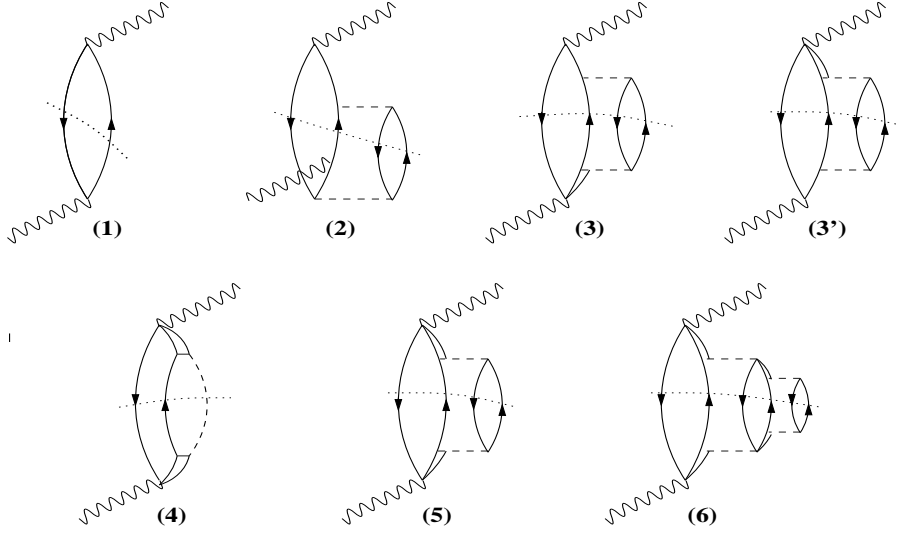}
\caption{\label{fig:1} Feynman graphs of the partial polarization 
propagators: $NN$ quasi-elastic (1), $NN$ (2p-2h) (2), $N\Delta$ (2p-2h) (3), 
$\Delta N$ (2p-2h) (3'), $\Delta\Delta$ ($\pi N$) (4), $\Delta\Delta$ (2p-2h) 
(5), $\Delta\Delta$ (3p-3h) (6). The wiggled lines represent the external probe, the full lines correspond to 
the propagation of a nucleon (or a hole), the double lines to the propagation of a Delta 
and the dashed lines to an effective interaction between 
nucleons and/or Deltas. The dotted lines show which particles are placed on-shell.}
\end{center} 
\end{figure}

The presence of the spin-isospin longitudinal coupling is 
a distinct feature of neutrino interaction as compared to inelastic electron scattering. For instance coherent pion production, 
present in $\nu$ interactions is partly suppressed in $(e,e')$ scattering due to the purely transverse spin coupling of the exchanged photon. 
Inclusive electron scattering is nevertheless  useful as a test for the transverse response 
\cite{Leitner:2008ue}. 
%In Section \ref{subs_qe} we give a test of this response.
The  response functions are related to the imaginary part of the corresponding full polarization propagators
\begin{equation} \label{eq:6}
R (\omega,q) = -\frac{\mathcal{V}}{\pi} \, \mathrm{Im}[ \Pi(\omega,q,q) ],
\end{equation}
where $\mathcal{V}$ is the nuclear volume such that $\mathcal{V}\rho=A$.
They are calculated within a RPA (random phase) ring approximation starting from ``bare'' propagators 
(meaning that the nuclear correlations are switched off). 
The word bare here does not imply that the corresponding response is free of many-body effects, as described in the following. 
The ``bare'' polarization propagator is illustrated by some of its components in Fig. \ref{fig:1} where the wiggled lines represent the external probe, 
the full lines correspond to the propagation of a nucleon (or a hole), the double lines to the propagation of a Delta and the dashed lines to an effective interaction between 
nucleons and/or Deltas.

The dotted lines in Fig. \ref{fig:1} indicate, in each of the channels introduced previously ($NN, N\Delta$ or $\Delta\Delta$), 
which intermediate state is placed on-shell. It follows that the bare response is the sum of the following partial components: 
%\begin{enumerate}       
%\item $ NN $~:~quasi-elastic (as described by the standard Lindhard function), 
%\item $ NN $~: \textit{2p-2h}, 
%\item $ N\Delta $  and $ 3^\prime $ $ \Delta N $~: \textit{2p-2h}, 
%\item $ \Delta\Delta  $~: $\pi \, N $,
%\item $ \Delta\Delta $~: \textit{2p-2h},
%\item $ \Delta\Delta $~: \textit{3p-3h}.
%\end{enumerate}
$ NN $~:~quasi-elastic (as described by the standard Lindhard function), $ NN $~: $2p-2h$, 
$ N\Delta $  and   $ \Delta N $~: $2p-2h$, 
$ \Delta\Delta  $~: $\pi \, N $, $ \Delta\Delta $~: $2p-2h$, 
$ \Delta\Delta $~: $3p-3h$.

Notice that the graphs shown in Fig. \ref{fig:1} do not exhaust all the possibilities for the bare propagator. 
For instance the distortion of the pion emitted by the $\Delta$ is not explicitly shown, although it will be included 
in our evaluation through the modification of the $\Delta$ width in the nuclear medium. But the type of final states 
that we consider is limited to the previous list. Thus, in the bare case, through the introduction of 
the partial polarization propagators illustrated by the Feynman graphs of  Fig. \ref{fig:1}, 
the inclusive expression of Eq. (\ref{SIGMANUENU}) provides an access to the exclusive ones, with  specific final states.

For the actual evaluation of the bare response, {\it i.e.,} the imaginary piece of the bare propagator, 
some of the 
graphs of Fig. \ref{fig:1} amount to a modification of the Delta width in the medium. 
We take into account this modification through the parameterization of the in-medium Delta width of 
Oset and Salcedo \cite{Oset:1987re}, which leads to a good description of pion-nuclear reactions. 
The authors split the $\Delta$ width  into different decay channel contributions~: 
the $ \Delta \longrightarrow \pi \, N $, which is modified by the Pauli blocking of 
the nucleon and the distortion of the pion. Moreover, in the nuclear medium, new decay channels are possible: 
the two-body ($2p-2h$) and three-body ($3p-3h$) absorption channels which they also incorporate. 
They  give a parametrization for the inclusion of these effects, both in the case of pion interaction with nuclei and 
for the photo-production process. 
We have used their parametrization in spite of the fact that 
in neutrino interaction the intermediate boson  has a space-like character.
An explicit evaluation
 of the corresponding contributions in the kinematical situation of neutrino scattering is desirable. 
There exist also other  $2p-2h$ contributions which are not reducible to a modification of the Delta width. 
We include them, as in the work of Marteau \cite{Marteau:1999kt}, 
following the method of Delorme and Guichon \cite{DELGUICH} 
who perform an extrapolation of the  
calculations of Ref. \cite{Shimizu:1980kb}  on the $2p-2h$ absorption of pions at threshold. For the last 
contribution only the imaginary part of the corresponding propagator is incorporated. 
The explicit expressions are given in Appendix \ref{appendix_bare}.
It turns out that for 
neutrino interaction it is the dominant contribution to the $2p-2h$ final state channel, as will be
 illustrated later. 
This  piece of the cross-section is subject to some uncertainty as this parametrization  
has not been constrained by specific experimental tests. This point will be discussed in more detail in 
Secs. \ref{subs_qe} and \ref{comp_qe}.

\begin{figure}
\begin{center}
\includegraphics[width=12cm,height=7cm]{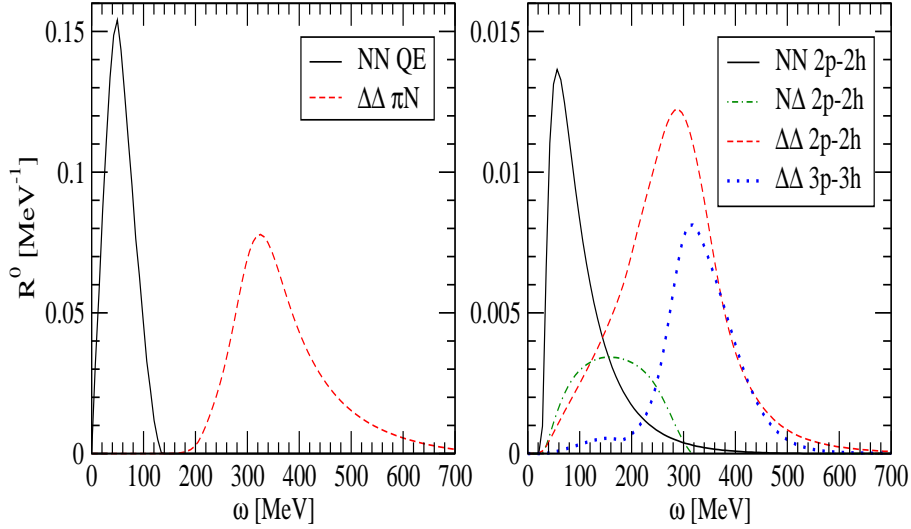}
\caption{Bare response for $^{12}$C at $q=300$ $\textrm{MeV}/c$ as a function of 
the energy transfer with its different components, quasi-elastic and pion emission (left panel), 
$2p-2h$ and $3p-3h$ (right panel).} 
\label{RESPONSE_BARE}
\end{center}
\end{figure}

The ``bare'' polarization propagator is density dependent. 
In a finite system, $\Pi^0(\omega,\bf{q},\bf{q}')$, it is non-diagonal in momentum space. 
In order to account for the finite size effects we evaluate it in a semi-classical approximation where it can be cast in the form 
\begin{equation} \label{eq:laktineh}
\Pi^0(\omega,\gvec{q},\gvec{q'}) = \int\, d \gvec{r} \,e^{-i(\gvec{q}-\gvec{q'}) \cdot \gvec{r}} \,
\Pi^0\left(\omega,\frac{1}{2}\left(\gvec{q}+\gvec{q'}\right),\gvec{r}\right).
\end{equation}
In practice we use a local density approximation,
\begin{equation}
\Pi^0\left(\omega,\frac{\gvec{q}+\gvec{q}'}{2},\gvec{r}\right)=
\Pi^0_{k_F(r)}\left(\omega,\frac{\gvec{q}+\gvec{q}'}{2}\right), 
\end{equation}
where the local Fermi momentum $ k_F(r) $ is related to the experimental nuclear density through~: 
$ k_F(r) = ( 3/2 \, \pi^2 \, \rho(r) )^{1/3} $.  
The density profiles of the various nuclei considered are taken from the 
Sum-of-Gaussians nuclear charge density distribution parameters according to Ref. \cite{De Jager:1987qc}.
The corresponding bare response for $^{12}$C at $q=300$ $\textrm{MeV}/c$ as a function of 
the energy transfer is illustrated in Fig.\ref{RESPONSE_BARE} with its different components, quasi-elastic, pion emission, $2p-2h$ and $3p-3h$. 
In all figures the responses incorporate the multiplicative spin-isospin factor.

Turning to the Random Phase Approximation,
as the semi-classical approximation is not suited to evaluate the collective effects, 
we have used the previous bare polarization propagator $ \Pi^0 $  as an input 
in a full quantum mechanical resolution of the RPA equations in the ring approximation. 
The introduction of the RPA correlations amounts to solving integral equations which have the generic form:
\begin{equation} \label{eq:4}
\Pi = \Pi^0 + \Pi^0 \, V \Pi,
\end{equation}
where $ V $ denotes the effective interaction between \textit{particle-hole} 
excitations. Its diagrammatic representation is given in Fig.\ref{RPA}. 
Some detailed expressions are given in Appendix \ref{appendix_rpa}.
In the spin-isospin channel the RPA equations couple the $ L $ and $T $ or the $ N$ and $ \Delta $ 
components of the polarization propagators. The effective interaction relevant in the isospin 
and spin-isospin channels is the crucial ingredient for determining the importance of the RPA effects. 
We use the parametrization in terms of $ \pi ,  \rho \ $ and contact pieces:

\begin{eqnarray} \label{eq:INTERACTION}
V_{NN} & = & ( f^\prime \, + \, V_\pi \, + \, V_\rho \, + \, V_{g^\prime} )  \,\, 
\gvec{\tau}_1 . \gvec{\tau}_2 \nonumber \\
V_{N \Delta} & = & ( V_\pi \, + \, V_\rho \, + \, V_{g^\prime} )  \,\, 
\gvec{\tau}_1 . \gvec{T}^\dagger_2 \nonumber \\  
V_{\Delta N} & = & ( V_\pi \, + \, V_\rho \, + \, V_{g^\prime} )  \,\, 
\gvec{T}_1 . \gvec{\tau}_2 \nonumber \\  
V_{\Delta\Delta} & = & ( V_\pi \, + \, V_\rho \, + \, V_{g^\prime} )  \,\, 
\gvec{T}_1 . \gvec{T}^\dagger_2 . 
\end{eqnarray}
For instance, in the $ NN $ case one has :
\begin{eqnarray}
V_\pi & = &  \left(\frac{g_r}{2 M_N}\right)^{2}\, F_\pi^2 \,\, \frac{\gvec{q}^2}{\omega^2 - \gvec{q}^2 - m_\pi^2} \,\,
 \gvec{\sigma}_1 . \widehat{q} \,\, \gvec{\sigma}_2 . \widehat{q} \nonumber \\
V_\rho & = &\left(\frac{g_r}{2 M_N}\right)^{2}\,C_\rho\, F_\rho^2 \,\, \, \frac{\gvec{q}^2}{\omega^2 - \gvec{q}^2 - m_\rho^2}  \,\,
 \gvec{\sigma}_1 \times \widehat{q} \,\, \gvec{\sigma}_2 \times \widehat{q} 
\nonumber \\
V_{g^\prime} & = &\left(\frac{g_r}{2 M_N}\right)^{2}\, F_\pi^2 \,\,\, g^\prime \,\,\, \gvec{\sigma}_1 . \gvec{\sigma}_2, 
\end{eqnarray}
where $g'$ is the Landau-Migdal parameter and $C_\rho=1.5$. Here $ F_\pi(q) =(\Lambda^{2}_{\pi}- m^{2}_{\pi})/(\Lambda^{2}_{\pi} - q^2)$ and 
$ F_\rho(q) =(\Lambda^{2}_{\rho}- m^{2}_{\rho})/(\Lambda^{2}_{\rho} - q^2)$ are the pion-nucleon and rho-nucleon form factors, 
with $\Lambda_{\pi}$ = 1 GeV and $\Lambda_{\rho}$ = 1.5 GeV. 
For the Landau-Migdal parameter $f^\prime$, we take  $f^\prime= 0.6 $.
 As for the spin-isospin  parameters  $g^\prime $ we use the information of the spin-isospin phenomenology \cite{IST06}, 
with a consensus for a larger value of $g'_{NN}=0.7$; for the other parameters we take $g'_{N\Delta}=g'_{\Delta\Delta}=0.5$.   
 
\begin{figure}
\begin{center}
\includegraphics[width=3cm]{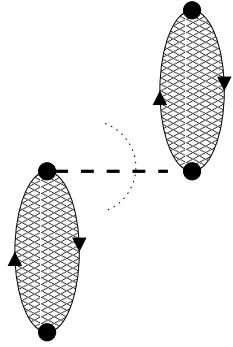}
\caption{Diagrammatic representation of the coherent process. The dotted line indicates that pion is placed on-shell.}
\label{RPA_coherent}
\end{center}
\end{figure}

The separation between the specific channels is less straightforward in the RPA case than in the bare one. Indications can be obtained with the following method, 
introduced in Ref. \cite{Marteau:1999kt}. The imaginary part of $\Pi$ can be written (again generically) as~:
\begin{equation} 
\mathrm{Im} \Pi = \left| \Pi \right|^2 \, \mathrm{Im} V \,+  \left| 1 + \Pi \, V \right|^2 \, \mathrm{Im}\Pi^0.  \,  \label{SEPAR}
\end{equation}
It separates into two terms. The first term on the r.h.s. of Eq.(\ref{SEPAR}), $ | \Pi |^2 \, \mathrm{Im} V $, is absent when the effective interaction is switched off. In the domain of energy considered it is the imaginary part of the pion exchange potential $V_\pi$
which plays the major role. This process thus represents the coherent pion production, \textit{i.e.}, 
the emission of an on-shell pion, the nucleus remaining 
in its ground state. This 
is illustrated in  Fig. \ref{RPA_coherent}, in which the hatched rings represents the RPA polarization propagator.  
The second  term on the r.h.s. of Eq. (\ref{SEPAR}), proportional to the bare polarization propagator $\mathrm{Im}\Pi^0$, 
reflects the type of final state already mentioned for the imaginary part of $\Pi_0$: $NN, \pi N,....$. 
The factor in front, $ | 1 + \Pi \, V |^2$, embodies the modification of the exclusive bare responses by 
the collective effects. We point out however that final state interactions are not incorporated in this description. 
For instance a pion produced in the decay of the Delta resonance can be absorbed on its way out leading to  a multi-nucleon emission process. 
Thus the second term in Eq. (\ref{SEPAR}) 
is adequate for  the sum of the incoherent pion production and the multi-nucleon knock-out channels, but not for each channel individually. 
The separation between these two channels from  the type of final state 
is approximate for light nuclei such as $^{12}$C. In heavier nuclei it overestimates the incoherent pion channel, 
underestimating the multi-nucleon one. We will illustrate this fact in the scattering of physical pions.

Having established the formalism, we are now ready to evaluate the  cross-sections in the various partial 
channels. In the actual numerical calculation we have limited the energy transfer to $\omega=1$ GeV as 
our approach becomes insufficient for a larger energy transfer. 
The center-of-mass correction for the 
$\pi$-$N$ system 
$q_{\textrm{CM}}=\frac{q}{1+\omega/M}$ \cite{Ericson:1988gk}
is made by dividing the bare responses by a factor $r^2=(1+\omega/M)^2$. 
The components of the neutrino cross-section that does not involve the momentum $q$ at the two ends of the RPA chain are 
obtained by an overall multiplication by the factor $r^2$. Interference terms with one momentum are multiplied by $r$.

\begin{figure}
\begin{center}
\includegraphics[width=6cm]{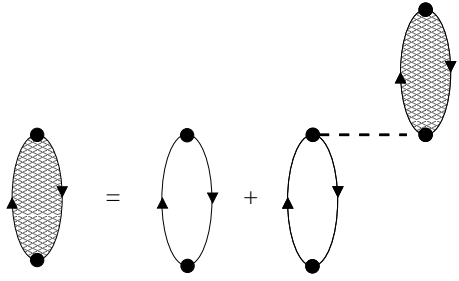}
\caption{Diagrammatic representation of the RPA polarization propagator. The white bubble is the free p-h propagator 
while the black is the full RPA one.} 
\label{RPA}
\end{center}
\end{figure}

%%%%%%%%%%%%%%%%%%%%%%%%%%%%%%%%%%%%%%%%%%%%%%%%%%%%%%%%%%%%%%%%%%%%%%%%%
\section{Results}
%%%%%%%%%%%%%%%%%%%%%%%%%%%%%%%%%%%%%%%%%%%%%%%%%%%%%%%%%%%%%%%%%%%%%%%%%%%%

%%%%%%%%%%%%%%%%%%%%%%%%%%%%%%%%%%%%%%%%%%%%%%%%%%%%%%%%%%%%%%%%%%%%%%%%%
\subsection{Coherent cross-section}
\begin{figure}
\begin{center}
\includegraphics[width=12cm,height=7cm]{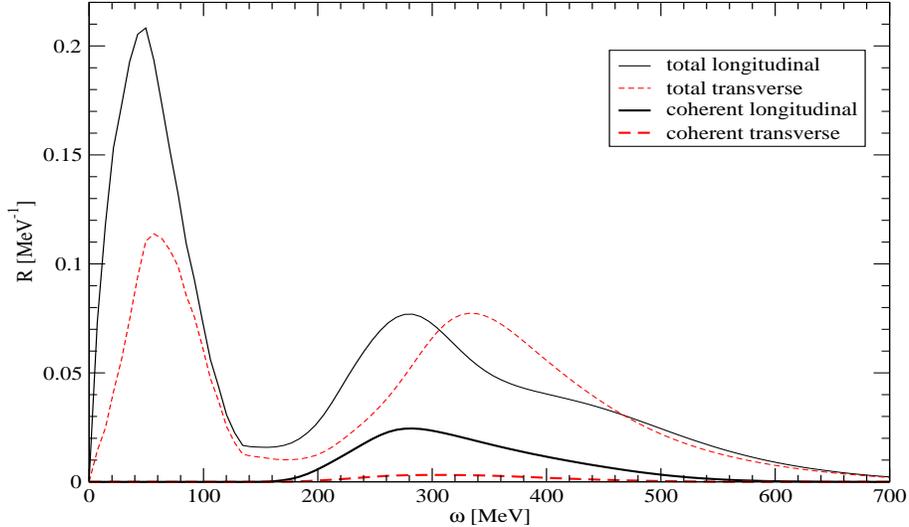}
\caption{Longitudinal and transverse total responses of $^{12}$C at fixed $q=300$ MeV/$c$ as a function of $\omega$. 
The coherent part of the responses is also shown.} 
\label{RESPONSE_TOT_COH}
\end{center}
\end{figure}

%%%%%%%%%%%%%%%%%%%%%%%%%%%%%%%%%%%%%%%%%%%%%%%%%%%%%%%%%%%%%%%%%%%%%%%%%%%%
Several types of responses enter the total neutrino cross-section, isovector, spin-isospin: transverse or longitudinal. 
The last quantity is naturally associated with the coherent process, since it has the same coupling as the pion. 
The production by a 
transverse spin coupling requires a transverse-longitudinal conversion which is partly suppressed. 
This difference is illustrated in Fig. \ref{RESPONSE_TOT_COH} where the total responses, longitudinal and transverse, 
of $^{12}$C are displayed as a function of the energy transferred to the nuclear system for 
a fixed three-momentum q= 300 MeV$/c$. 
The coherent component, 
 much larger in the longitudinal case, is also shown.

\begin{figure}
\begin{center}
\includegraphics[width=12cm,height=7cm]{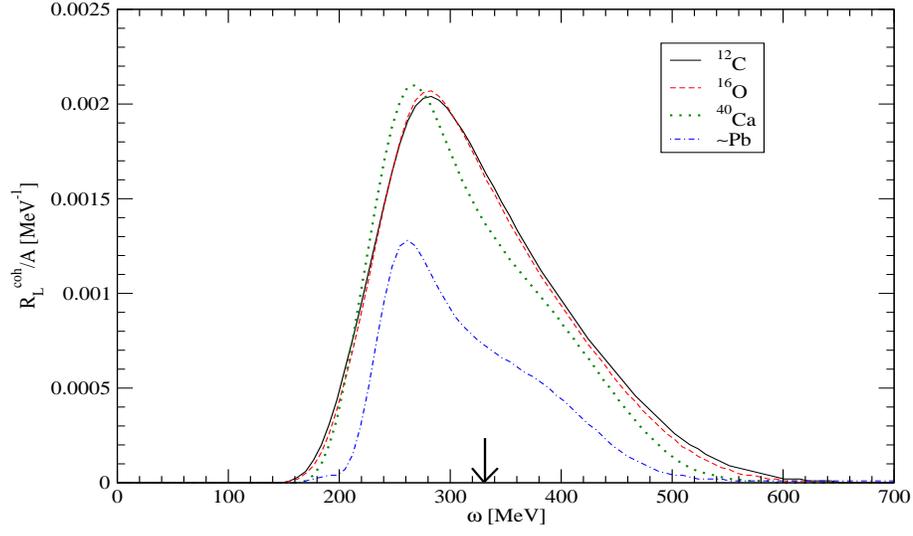}
\caption{Evolution with the mass number of the coherent longitudinal response per nucleon 
at fixed $q=300$ MeV/$c$ as a function of $\omega$. The arrow indicates the energy for on-shell pion.} 
\label{RESPONSE_COH_NNUCLEI}
\end{center}
\end{figure} 

Figure \ref{RESPONSE_COH_NNUCLEI} illustrates the evolution with the nuclear size of the coherent part of the longitudinal response {\bf \textit{per nucleon}} as a function of the energy at fixed momentum for some nuclei, 
$^{12}$C, $ ^{16}$O, $^{40}$Ca 
and also for a fictitious piece of isospin symmetric nuclear matter with the density profile of lead.  
Two features emerge, the first one is that its magnitude decreases in ``lead'', as expected~: 
the coherent response per nucleon vanishes in nuclear matter when the polarization propagators become diagonal in momentum space. 
The second is that the coherent response is not peaked at the energy $\omega_{\pi}= (q^2 +m_{\pi}^2)^{1/2}$ 
where the mismatch between the incident energy and that of the physical outgoing pion  is smallest. 
Instead it is reshaped by the collective features of the longitudinal response with the appearance of two 
collective branches on each side of the pion line. 
This is more apparent in the case of the (fictitious) lead.

\begin{figure} 
\begin{center}
  \includegraphics[width=12cm,height=8cm]{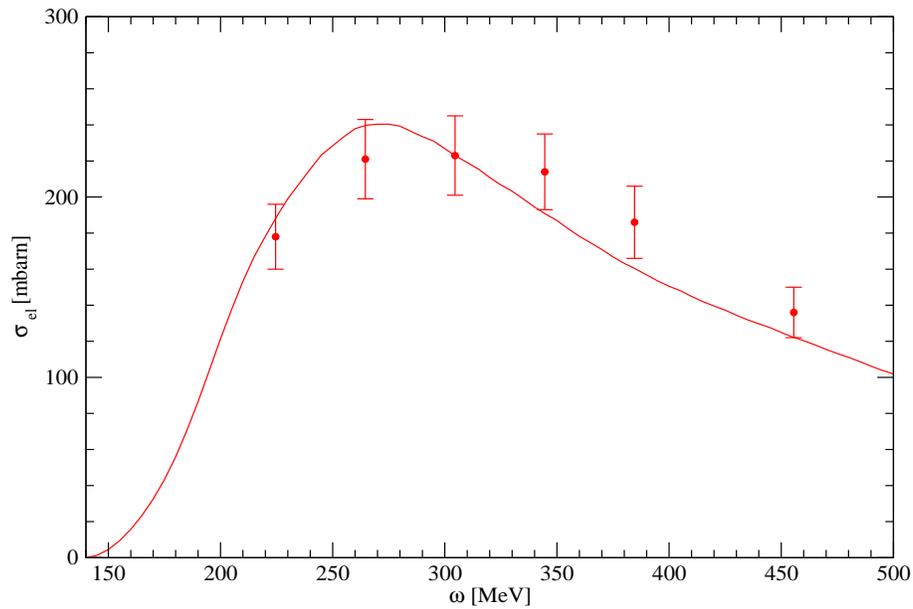}
{\caption{$\pi$-$^{12}$C elastic cross-section as a function of pion energy.}\label{fig_pi_c12_el}}
%\label{fig_pi_comp}
\end{center}
\end{figure}

\begin{figure} 
\begin{center}
  \includegraphics[width=12cm,height=8cm]{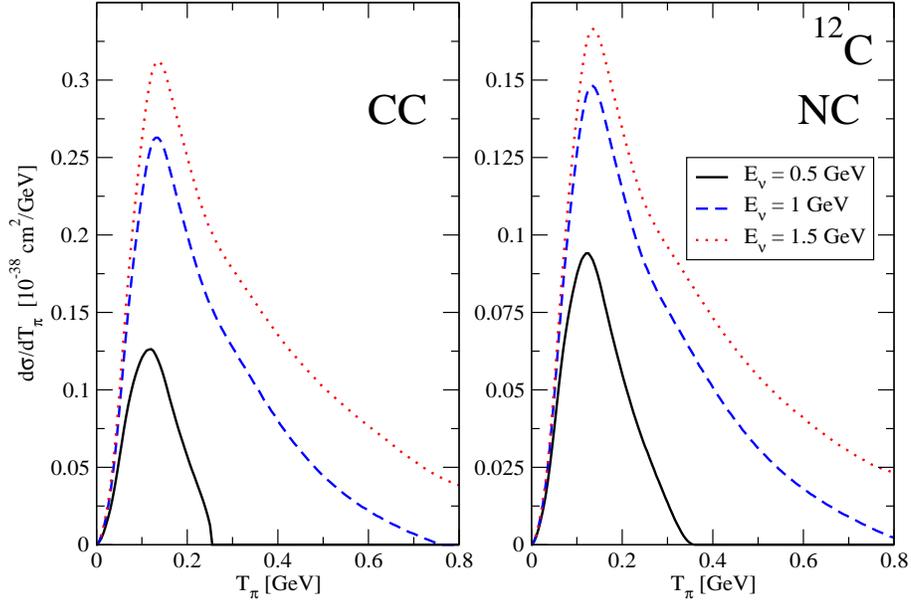}
{\caption{\label{fig_coh_diff}
Charged and neutral current coherent pion production 
differential cross-section off $^{12}$C versus pion kinetic energy for 
several $\nu_\mu$ energies.}}
%\label{fig_coh_diff}
\end{center}
\end{figure}
\begin{figure} 
\begin{center}
\includegraphics[width=12cm,height=8cm]{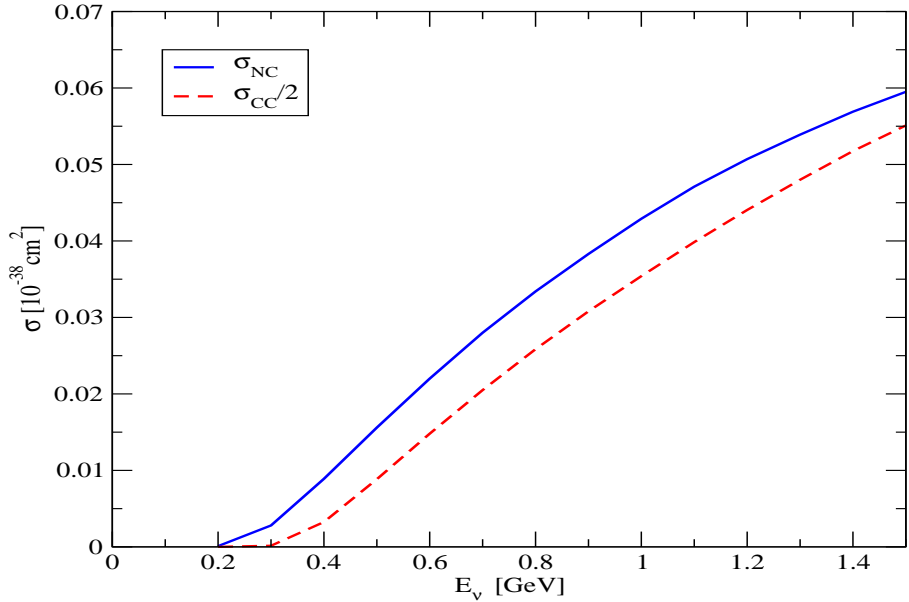}
{\caption{Total CC (divided by 2) and NC $\nu_\mu$-induced coherent pion production 
cross-sections in $^{12}$C as a function of neutrino energy.}\label{fig_coh_tot}}
%\label{fig_coh_tot}
\end{center}
\end{figure}

As a test of our description of the coherent responses we have investigated the elastic scattering of pions on
 nuclei in the Delta region, related to the coherent part of the spin-isospin longitudinal response through~:

\begin{equation} \label{ELASTIQUE}
\sigma^{elas}(\omega) =  \left(\frac{g_r}{2M_N}\right)^2 {\pi}q_{\pi}\, 
R_L^{coh}(\omega,q_{\pi}) 
\end{equation}
where $q_{\pi}^2=\omega^2-m_{\pi}^2$ and $R_L^{coh}$ refers to the coherent part of the longitudinal response.
The resulting cross-section in the case of $^{12}$C is shown in Fig.\ref{fig_pi_c12_el}  together with the 
experimental points from Ref.\cite{Ashery:1981tq}. 
The agreement with data is satisfactory. A similar accuracy can be expected for the coherent response which enters 
the neutrino cross-section, at least in the energy region for the produced pion where 
we have tested our model (\textit{i.e.}, between $\omega \simeq 220$ and $\simeq 450 MeV$). 
The elastic cross-section which depends on the longitudinal response is particularly sensitive to 
collective effects in this channel known to be important. The replacement of the bare response by the 
RPA one leads to a different energy behavior, the collective effects producing a  softening of the response, 
characteristic of the collective nature of the longitudinal channel.\\
Figure \ref{fig_coh_diff} displays our evaluations of the neutrino coherent cross-section on $^{12}$C as a function 
of the pion kinetic energy, both for charged and neutral current, 
for several neutrino incident energies. The resulting total coherent cross-sections are displayed in Fig.\ref{fig_coh_tot}.\\ 
The suppression of the meson exchange correction in the time component of the axial current, $G_A^*\to G_A$, 
produces a moderate $\simeq 10 \%$ increase of the cross-section.\\ 
The data available on the coherent production by neutrino concern its ratio to the total cross-section and to the total pion production. 
We will then postpone the comparison with experimental data after the discussion of the various other channels.

\subsubsection{Adler's theorem}
\begin{figure} 
\begin{center}
  \includegraphics[width=12cm,height=8cm]{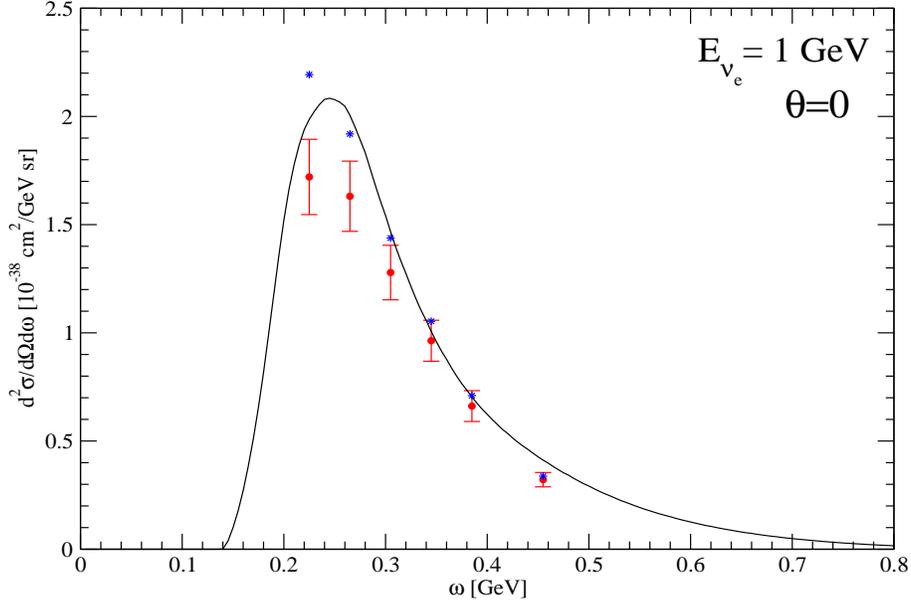}
\caption{$\nu_e$-$^{12}$C coherent cross-section in the forward direction.
Continuous line: our result. Circles: deduced, according to Adler's relation of Eq. (\ref{eq:coh_adler}), 
 from the experimental values for the elastic cross-section taken from Ref. 
\cite{Ashery:1981tq}.
Stars: introducing into the r.h.s. of Eq. (\ref{eq:coh_adler}) the multiplicative factor $\frac{\omega}{q_{\pi}}$.}
\label{test_adler}
\end{center}
\end{figure}

In the forward direction where $q=\omega$ and for vanishing lepton mass, only the spin longitudinal response contribution survives. 
As it also enters in pion scattering, it is possible to relate the forward neutrino cross-section to 
the cross-section of physical pions, 
apart from a difference in kinematics: $q=\omega$ (soft pions) for neutrinos, instead 
of $q=q_{\pi}=\sqrt{\omega^2-m_{\pi}^2}$ for physical pions. This difference becomes less relevant at large energies.
This is the content of Adler's theorem \cite{Adler:1964yx}. The coherent channel, which is completely dominated by the 
longitudinal response, offers the best application of this theorem, while 
for the other channels the transverse component, which bears no relation to pion scattering, quickly takes over as soon as one moves away from the forward direction. 
This theorem has been used in the approach of Refs. \cite{Paschos:2005km} \cite{Berger:2008xs} \cite{Paschos:2009ag} 
to evaluate the coherent neutrino-nucleus cross-section.
This is not our aim here. We want to illustrate the link between the forward 
direction coherent neutrino cross-section and the elastic pion-nucleus one.
For the coherent cross-section Adler's relation writes
\begin{equation}
\label{eq:coh_adler} 
\left(\frac{\partial^2\sigma}{\partial\Omega \,\partial \omega}\right)_{\theta=0}^{\textrm{coh}} =  \frac{G_F^2 \, 
\cos^2\theta_c }{\pi^3}f_{\pi}^2 \frac{E_{\nu}-\omega}{\omega}\sigma^{elas}(\omega),
\end {equation}
where $f_{\pi}=93.2$ MeV is the neutral pion decay constant.
Introducing the experimental values for the elastic cross-section taken from Ref.\cite{Ashery:1981tq}
we obtain the points shown in Fig. \ref{test_adler} together with our predicted curve. The agreement is 
rather good. It deteriorates at small energies when the kinematical difference between 
soft and physical pions becomes substantial. A natural correction can be performed 
with the introduction into the r.h.s. of Eq.(\ref{eq:coh_adler}) of a multiplicative factor $\frac{\omega}{q_{\pi}}$ as 
suggested by the relation of Eq.(\ref{ELASTIQUE}) between $R_L$ and $\sigma^{elas.}$. The corresponding 
corrected points are also shown in Fig. \ref{test_adler} extending somewhat the region of agreement. 
The use of the Adler relation 
becomes problematic at energies near threshold. For small neutrino energy ($E_{\nu}<0.5$ GeV) this region has more weight 
in the total coherent cross-section.

The Adler relation thus provides a good test for our evaluation of coherent neutrino cross-section in the forward direction. 
We believe that the extrapolation to the non-forward direction 
as performed in our model should be under control.

%%%%%%%%%%%%%%%%%%%%%%%%%%%%%%%%%%%%%%%%%%%%%%%%%%%%%%%%%%%%
\subsection{Pion-nucleus cross-sections}
%%%%%%%%%%%%%%%%%%%%%%%%%%%%%%%%%%%%%%%%%%%%%%%%%%%%%%%%%%%%%%

\begin{figure} 
\begin{center}
  \includegraphics[width=12cm,height=8cm]{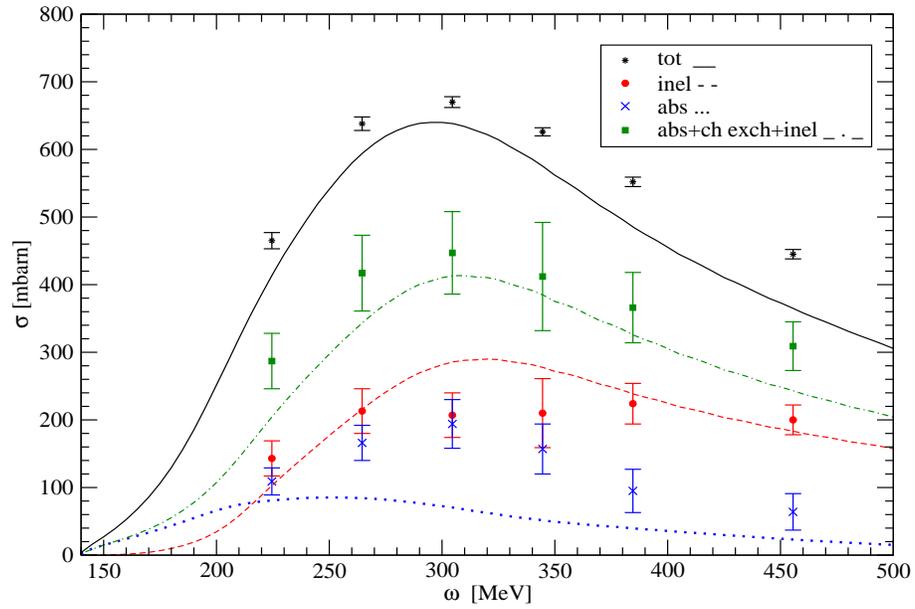}
{\caption{Partial and total $\pi$ - $^{12}$C cross-sections.}\label{fig_pi_comp}}
%\label{fig_pi_comp}
\end{center}
\end{figure}

The various partial cross-sections for physical pions on nuclei constitute a precious piece of information. 
Elastic cross-section has already been introduced 
as a test for the coherent cross-section. 
The total cross-section for pions on the nuclei is given by an expression similar to Eq.(\ref{ELASTIQUE}) 
with the full polarization propagator replacing the coherent piece
\begin{equation} \label{TOTAL}
\sigma^{tot}(\omega) = \left(\frac{g_r}{2M_N}\right)^2 {\pi}q_{\pi} 
\,R_L(\omega,q_{\pi}). 
\end{equation}
The corresponding cross-section is displayed in Fig.\ref{fig_pi_comp} 
together with the experimental points. 
We will show that in the same way the inelastic cross-section 
provides some information on the incoherent pion production by neutrinos and the absorptive cross-section 
on the multi-nucleon channels. 
Figure \ref{fig_pi_comp} displays the various partial channels 
(but the elastic one, previously shown) which contribute to the $\pi^+$ cross-section on $^{12}$C, 
namely the inelastic pion scattering channel 
(which is the incoherent scattering with a $\pi^+$ in the final state) and the absorptive one. We also display 
the sum of the incoherent pion 
(including charge exchange) and true absorption (multi-nucleon channels) cross-sections. 
The experimental points are taken from Ashery \textit{et al.} \cite{Ashery:1981tq}. 
To reduce the clutter, we have not explicitly plotted the 
charge-exchange cross-section which, in our approach, 
is one fifth of the inelastic $\pi^+$ cross-section and is consistent with the experimental data. 
While the elastic cross-section was well reproduced, 
our approach overestimates the $\pi^+$ inelastic channel in the peak region and largely underestimates the absorptive channel. 
We attribute this deficiency to the absence of pion final state interaction 
as the pion can be reabsorbed on its way out the nucleus. It can also undergo charge exchange 
process but this is a smaller effect. 
As a counterpart the absorptive multi-nucleon production is  underestimated, as is apparent in Fig.\ref{fig_pi_comp}. 
The sum of the two channels is instead reasonably well reproduced in the peak region.\\
These limitations also affect the incoherent neutrino-nucleus cross-section but 
we stress that, in contradistinction, our description for the coherent channel 
automatically contains the final state interactions and no further correction is needed. 
The total neutrino cross-section is also obviously not affected. 
With the information on the pion energy spectrum 
in neutrino interactions (that our calculation does not provide) 
it would be possible to estimate at each energy an attenuation factor for 
the incoherent neutrino production from the difference between 
our calculation and inelastic data for physical pions. 
For instance, for $^{12}$C at $E_\nu=1$ GeV, a rough evaluation of the 
overall correction for the incoherent production cross-section
with the information on the pion spectrum \cite{dytman}
results in a moderate reduction  of $\simeq 15\%$. 
A similar attenuation was found in oxygen at $E_\nu=500$ MeV and $E_\nu=750$ MeV \cite{Singh:1998ha}.
A larger correction is obviously expected for calcium.

%%%%%%%%%%%%%%%%%%%%%%%%%%%%%%%%%%%%%%%%%%%%%%%%%%%%%%%%%%%%%
\subsection{Quasi-elastic and multi-nucleon channels}
\begin{figure}
\begin{center}
\includegraphics[width=3cm]{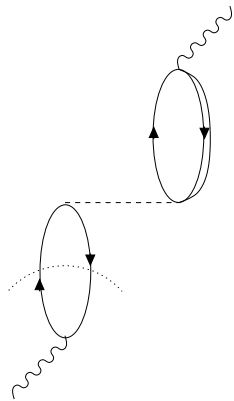}
\caption{Lowest order contribution of $R^{N\Delta}$ to the 
quasi-elastic channel.} 
\label{N_Delta_1order}
\end{center}
\end{figure}

%%%%%%%%%%%%%%%%%%%%%%%%%%%%%%%%%%%%%%%%%%%%%%%%%%%%%%%%%
\label{subs_qe}
The quasi-elastic (QE) channel corresponds to a single nucleon knock-out. 
In the quasi-elastic process the space-like character is  pronounced as 
the quasi-elastic peak occurs at $\omega\simeq{{\bf q}^2/(2M_N)}$, 
hence the distribution in $Q^2={\bf q}^2-\omega^2$ is rather broad 
\cite{:2007ru}.
At zero order only $R^{NN}$ contributes to this channel. 
In the RPA chain instead $R^{N\Delta}$ and 
$R^{\Delta \Delta}$ also participate. For instance the lowest order contribution of $R^{N\Delta}$ is illustrated in Fig. 
\ref{N_Delta_1order}.\\
In contrast to the coherent channel, the quasi-elastic one is totally dominated by the transverse response. 
 The longitudinal contribution is 
suppressed by a cancellation between the space and time components of the axial current, as observed by Marteau  
\cite{Marteau:1999kt} 
and shown in Appendix \ref{appendix_long_QE} for vanishing lepton mass and neglecting 
the Fermi momentum. Numerically its contribution is indeed very small. 
We have tested our semi-classical approximation on the bare QE $\nu_e$ - $^{12}$C cross-section 
through a comparison 
with the one obtained by Martini \textit{et al.} \cite{Martini:2007jw} in the continuum shell model 
where the mean field is produced by a 
Woods-Saxon well. 
Our result is very similar in shape and magnitude to the one of \cite{Martini:2007jw} but for a displacement 
in energy of $27$ MeV. 
This reflects the inclusion of the nucleon separation energy in the continuum shell model, 
which is ignored in our approximation.

\begin{figure}  
\begin{center}
  \includegraphics[width=12cm,height=8cm]{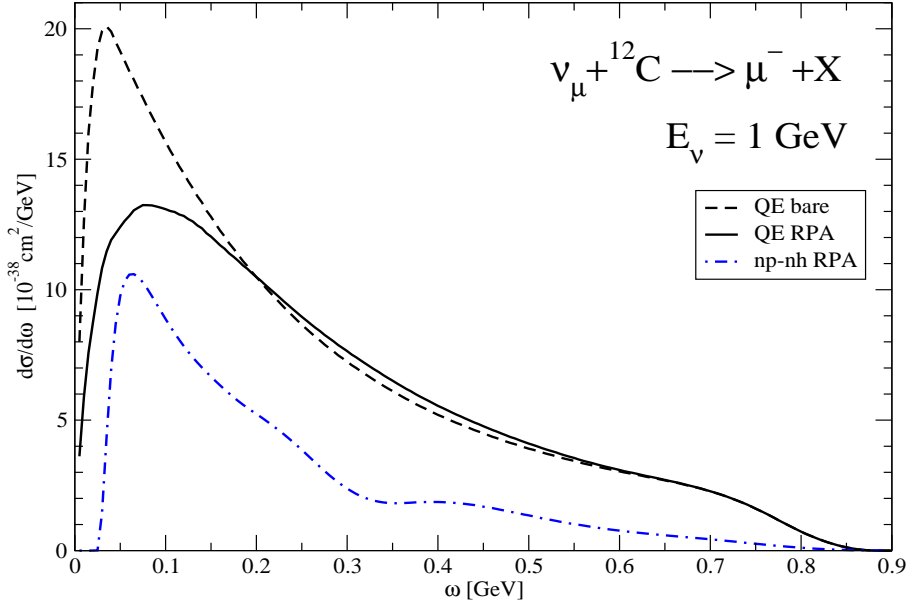}
\caption{Differential CC $\nu_{\mu}$ -- $^{12}$C cross-section versus the energy transfer for quasi-elastic process 
(bare and RPA) and multi-nucleon emission ($np-nh$).}
\label{fig_qe_np_c12}
\end{center}
\end{figure}

The quasi-elastic cross-section is displayed in Fig. \ref{fig_qe_np_c12} 
as a function of the energy transfer for neutrino energy $E_\nu=1$ GeV,
both in the bare case and in the RPA one. The RPA influence produces a reduction, 
as expected from the repulsive character of the particle-hole 
interaction, which prevails in the transverse channel. 
This reduction is mostly due to the interference term $R^{N\Delta}$ which is negative 
(Lorentz-Lorenz effect \cite{Ericson:1966fm}).

\begin{figure}
\begin{center}
  \includegraphics[width=12cm,height=8cm]{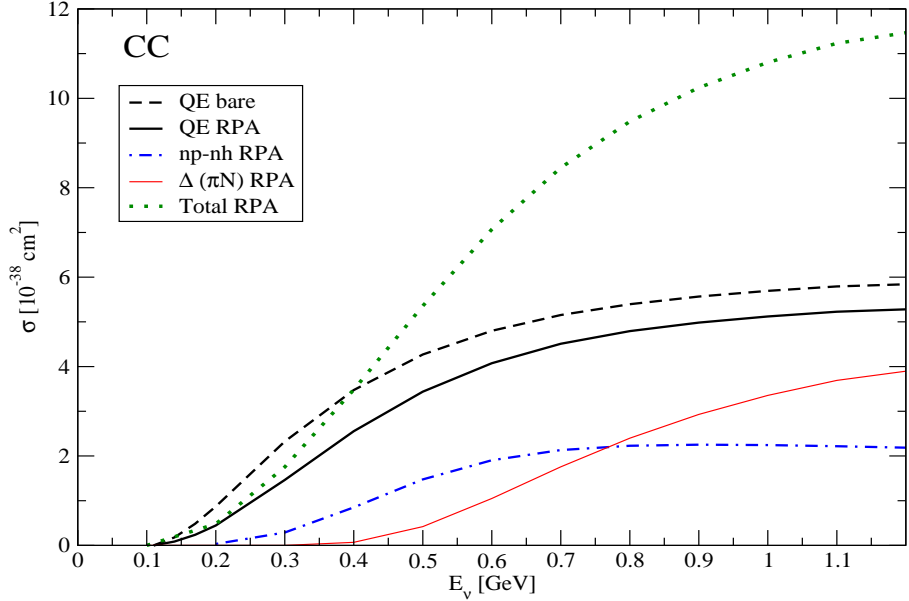}
\caption{CC $\nu_{\mu}$ -- $^{12}$C QE, multi-nucleon ($np-nh$), incoherent pion emission 
and total cross-section as a function of neutrino energy.}
\label{fig_cc_xtot}
\end{center}
\end{figure}

\begin{figure}
\begin{center}
  \includegraphics[width=12cm,height=8cm]{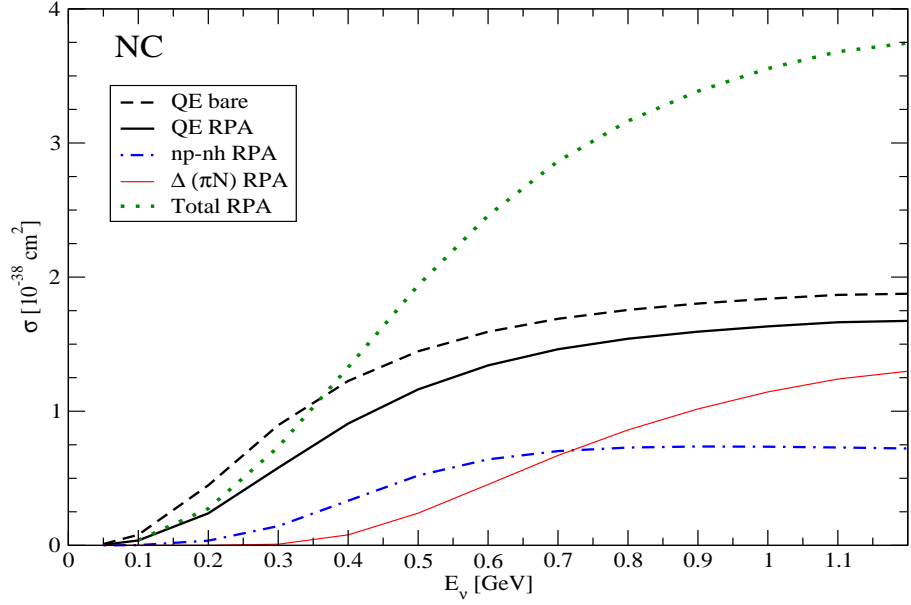}
\caption{NC $\nu_{\mu}$ -- $^{12}$C QE, multi-nucleon ($np-nh$), incoherent pion emission 
and total cross-section as a function of neutrino energy.}
\label{fig_nc_xtot}
\end{center}
\end{figure}

The total quasi-elastic charged current and neutral current cross-section are plotted 
in Figs.\ref{fig_cc_xtot} and \ref{fig_nc_xtot} as a function of the neutrino energy. 
In Figs. \ref{fig_qe_np_c12}, \ref{fig_cc_xtot} and \ref{fig_nc_xtot} 
we also display the sum of the two- and three-nucleon knock-out cross-sections, 
which represents a sizable fraction of the quasi-elastic one. 
Singling out the genuine quasi-elastic process requires the insurance that no more than one proton is ejected. 
This question will appear in the comparison with data.
Among the various 
contributions to the multi-nucleon channel the ones 
which do not reduce to a modification of the $\Delta$ width are dominant. 
The accumulation of $2p-2h$ strength at low energy is an artifact of the simplified extrapolation 
that we use in this channel. In Section \ref{comp_qe} this point is discussed in more detail 
and another method for the parametrization, with an explicit momentum dependence, is introduced. It modifies 
the $\omega$ dependence of $\frac{d \sigma}{d \omega}$, spreading the strength over a larger energy region, 
but does not substantially affect the energy integrated cross-section.

Coming now to the evolution of these channels between $^{12}$C and $^{40}$Ca 
we compare the corresponding RPA differential cross-sections per neutron 
for the two nuclei in Fig. \ref{c12_ca40_excl}.
One can see that the evolution of this quantity with the mass number is quite weak in the QE case. 
It is also weak in the multi-nucleon channel although it should increase faster with density than the quasi-elastic one. 
However, between a light system such as $^{12}$C and $^{40}$Ca the evolution is moderate. 
Only in the case of deuteron one expects the multi-nucleon knock-out 
to be appreciably smaller in view of the loose binding of the system.

\subsection{Incoherent pion emission}

\begin{figure}
\begin{center}
  \includegraphics[width=12cm,height=8cm]{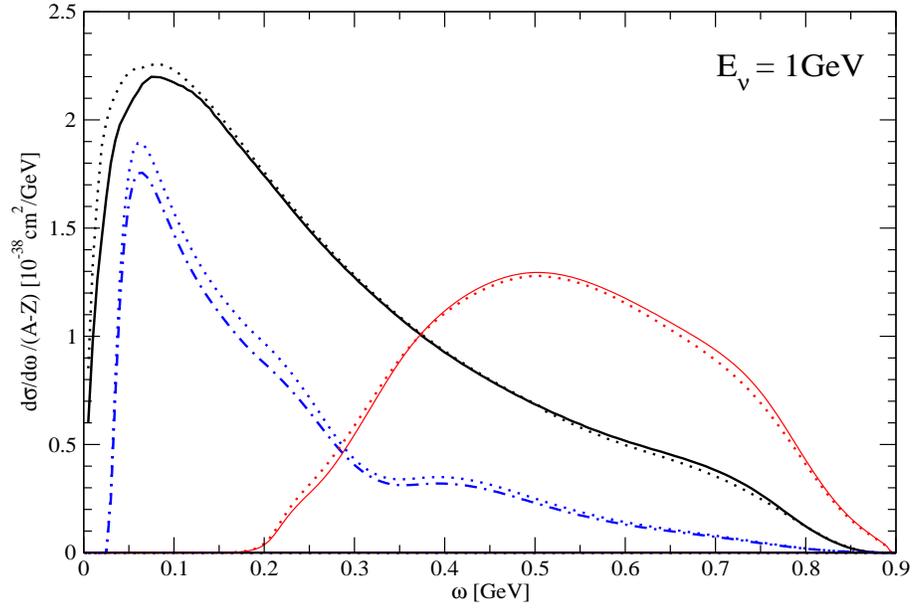}
\caption{RPA differential CC cross-sections per neutron in the different channels
for $^{12}$C (same convention-line as Figs.\ref{fig_qe_np_c12},\ref{fig_cc_xtot},\ref{fig_nc_xtot}), and $^{40}$Ca (dotted lines).}
\label{c12_ca40_excl}
\end{center}
\end{figure}

\begin{figure}
\begin{center}
  \includegraphics[width=12cm,height=8cm]{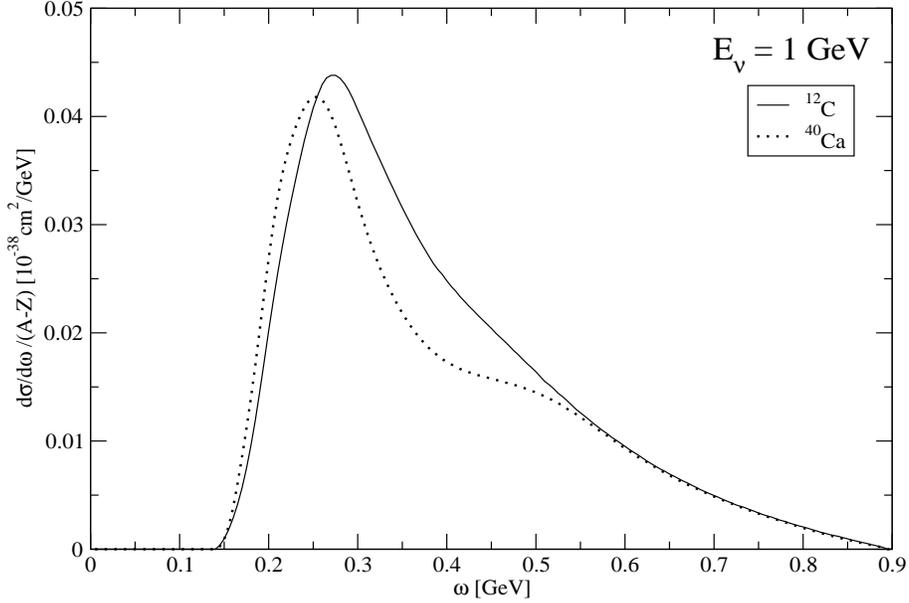}
\caption{Differential CC cross-sections per neutron in the coherent channel
for $^{12}$C (continuous line), and $^{40}$Ca (dotted line).}
\label{c12_ca40_coh}
\end{center}
\end{figure}

The pion arises from the pionic decay of the Delta leaving the nucleus in a $p-h$ excited state. 
For the nuclei that we consider this cross-section is much larger than the coherent one. 
As compared to a free nucleon the emission probability is already appreciably reduced 
in the bare case by the change in the Delta width.  
Moreover the RPA effects, which are moderate, also tend to a small reduction. 
The reduction due to the modification of the Delta width has a counterpart in 
the presence of a component of multi-nucleon knock-out. 
Charged current and neutral current cross-sections for 
incoherent pion emission for all possibles charges are represented in Fig.\ref{fig_cc_xtot} and \ref{fig_nc_xtot} 
as a function of neutrino energy. Moreover these figures summarize all previous results for the other channels 
and also give the total cross-sections.

On the other hand, Fig.\ref{c12_ca40_excl} 
compares the neutrino differential cross-section per neutron in the various channels as a function of the energy transfer, $\omega$,  
for the cases of $^{12}$C and $^{40}$Ca and for a neutrino energy $E_{\nu}= 1$ GeV. 
The two sets of curves are very similar. 
We can conclude that, at the level of our approximation, \textit{i.e.}, 
without final state interaction, it is possible to extrapolate smoothly from $^{12}$C to the region 
of $^{40}$Ar. Only the coherent cross-section presents a significant variation, illustrated in Fig.\ref{c12_ca40_coh}.

%%%%%%%%%%%%%%%%%%%%%%%%%%%%%%%%%%%%                    
\section{Comparison with data}
%%%%%%%%%%%%%%%%%%%%%%%%%%%%%%%%%%%%%%
\subsection{Coherent pion production}
%%%%%%%%%%%%%%%%%%%%%%%%%%%%%%%%%%%%%%%%
\begin{figure}
\begin{center}
  \includegraphics[width=12cm,height=8cm]{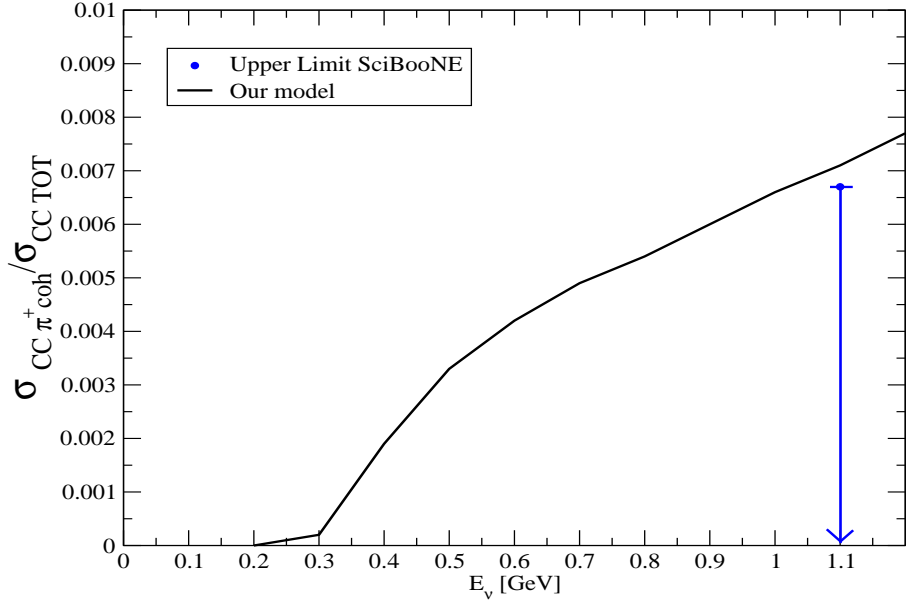}
\caption{Ratio of the $\nu_{\mu}$-induced charged current coherent
$\pi^+$ production to total cross-section as a function of neutrino energy.}
\label{fig_coh_ratio}
\end{center}
\end{figure}

Experimental data concern ratios between different cross-sections. 
The first indication of coherent pion production by neutral current was given by MiniBooNE 
\cite{AguilarArevalo:2008xs},
that found the ratio of coherent to total $\pi^0$ production to be 0.195 
$\pm$ 0.011 $\pm$ 0.025. In this experiment the neutrino 
flux is spread in energy with a  peak at $\simeq700 MeV$ 
\cite{AguilarArevalo:2008yp}. 
Our approach leads to a lower number, namely 0.06, which is difficult to reconcile with experimental data, 
a problem that other groups also face. 
It has been suggested in Ref.\cite{Amaro:2008hd} that 
MiniBooNE, which uses Rein-Sehgal model \cite{Rein:1982pf} 
for data analysis, possibly overestimates the $\pi^0$ coherent cross-section.
In a preliminary report \cite{Raaf} the experimental value given for this cross-section is
 $7.7 \pm 1.6\pm3.6~10^{-40}$ cm$^2$. Our result for this cross-section 
averaged on the MiniBooNE flux \cite{AguilarArevalo:2008yp}, 2.8 10$^{-40}$ cm$^{2}$,
is compatible with the experiment in view of 
the large experimental errors.

On the other hand 
for charged current, 
two experimental groups have given upper limits 
for the 
ratio of coherent pion production to 
the total cross-section.
The K2K collaboration gives a 
limit of $0.60~10^{-2}$ averaged over a neutrino flux 
with a mean energy of 1.3 GeV \cite{Hasegawa:2005td}. 
More recently, the SciBooNE collaboration found for the same quantity $0.67~10^{-2}$ 
at neutrino energy of 1.1 GeV \cite{Hiraide:2008eu} and $1.36~10^{-2}$ at neutrino energy of 2.2 GeV. 
We report in Fig.\ref{fig_coh_ratio} our prediction for this quantity. Since our approach is appropriate
for a limited neutrino energy range we keep in the comparison only the lowest energy SciBooNE point. 
Our curve is just compatible with the experimental bound.

\subsection{Total pion production}

\begin{figure}
\begin{center}
  \includegraphics[width=12cm,height=8cm]{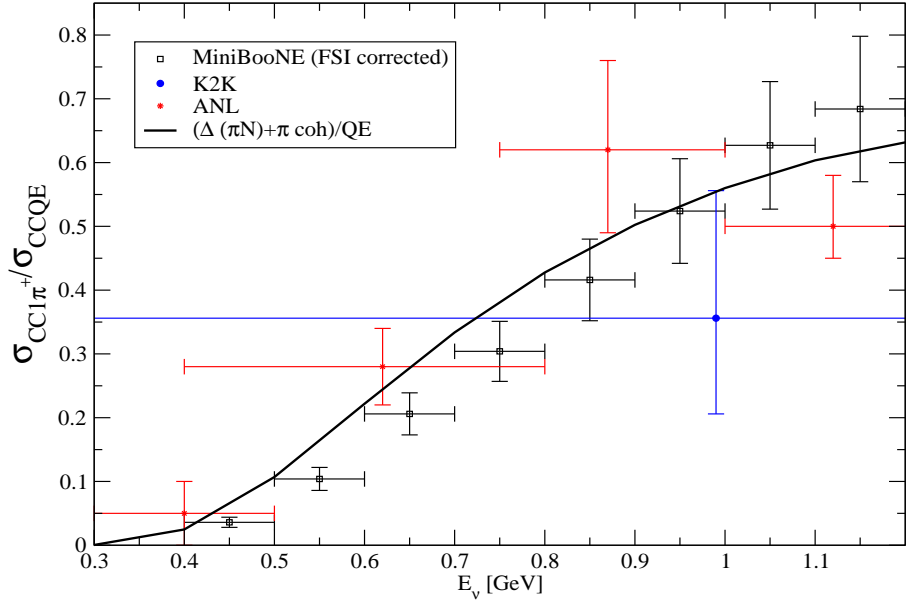}\\
\vskip 2.0 true cm
  \includegraphics[width=12cm,height=8cm]{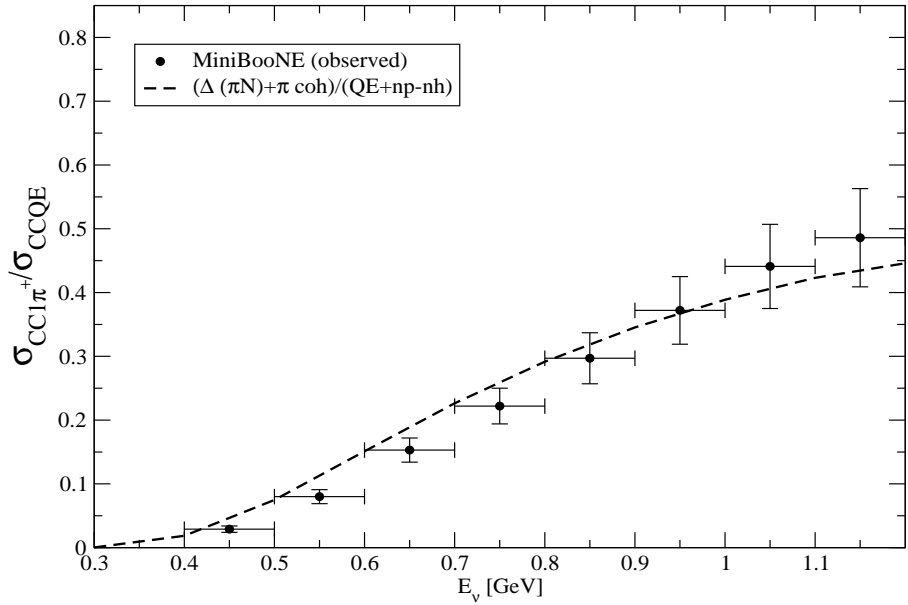}
{\caption{Ratio of the $\nu_{\mu}$-induced charged current one $\pi^+$ production to 
quasi-elastic cross-section as a function of neutrino energy.}\label{fig_ratio_pip}}
\end{center}
\end{figure}

Another measured quantity is the ratio of $\pi^+$ production to 
quasi-elastic cross-section for charged current. The MiniBooNE collaboration has used a 
CH$_2$ target. In order to compare with ANL \cite{Radecky:1981fn} and K2K \cite{:2008eaa}
data, they presented the results with an isoscalar rescaling correction \cite{AguilarArevalo:2009eb}. 
The issue of pion loss by final state interaction, which is not incorporated in our description, 
has also been taken into account by MiniBooNE 
who corrects data for this effect.
We can thus compare our $\pi^+$ over quasi-elastic ratio (solid line in the upper 
panel of Fig.\ref{fig_ratio_pip})
to the final-state-interaction-corrected MiniBooNE results. 
Our curve incorporates the small coherent cross-section; the incoherent pion one is multiplied by the isospin 
factor $5/6$ to single out $\pi^+$ contribution.
Our curve is fully compatible with experimental data.

As an additional information, MiniBooNE also gives a ratio more directly related to the measurements,
namely the ratio of pion-like events 
(defined as events with exactly one $\mu^-$ and one $\pi^+$ escaping the struck nucleus) 
and quasi-elastic signal (defined as those with one $\mu^-$ and no pions). 
In our language the last quantity represents the total $Np-Nh$ 
($N=1,2,3$, including the quasi-elastic for $N=1$) exclusive channel. 
We have compared this second experimental information to 
the ratio between our calculated pion production (which however ignores final state interactions) 
and our total $Np-Nh$ 
contribution to the total charged current neutrino cross-section 
(lower panel of Fig.\ref{fig_ratio_pip}). 
There is an appreciable difference between the two curves of Fig.\ref{fig_ratio_pip}:
the one in the lower panel is reduced due to 
a large $2p-2h$ component in the $Np-Nh$ cross-section, which increases the denominator.
The comparison with the experiment shows an 
agreement up to  $E_{\nu}\simeq 1.2$ GeV.
Final state interactions for the pion, which are not included, are expected to reduce our result
at the level of 15 $\%$, still maintaining an agreement.

A new result has been presented at NuInt09 by SciBooNE \cite{Kurimoto:2009yb}. 
It is the ratio of the total neutral current $\pi^0$ 
production cross-section to the total charged current cross-section at the mean neutrino neutrino energy 
of $1.16$ GeV. 
They obtain the preliminary value:
\begin{equation}
\frac{\sigma (NC~\pi_0)}{\sigma(CC_{TOT})}=(7.7 \pm 0.5(\textrm{stat.})^{+0.4}_{-0.5}(\textrm{sys.}))\cdot 10^{-2}.
\end{equation}
Our prediction for this quantity, including coherent contribution and a factor $2/3$ 
for NC incoherent pion production to single out $\pi^0$ contribution is: 
\begin{equation}
\frac{\sigma (NC~\pi_0)}{\sigma(CC_{TOT})}=7.9 \cdot 10^{-2},
\end{equation}
which fully agrees with data.

A general comment on the comparison with data: nearly all the ratios that have been discussed, 
except the final-state-interaction-corrected 
MiniBooNE result of the upper panel of Fig.\ref{fig_ratio_pip}, are sensitive to the presence of the $np-nh$ ($n=2,3$) 
component in the cross-section. 
Since the size magnitude is not so well tested, 
we can investigate what becomes the comparison with data 
in the extreme situation when we totally suppress this contribution.
For the last ratio discussed we obtain
\begin{equation}
\frac{\sigma (NC~\pi_0)}{(\sigma(CC_{TOT})-\sigma(CC_{np-nh}))}=9.8 \cdot 10^{-2},
\end{equation}
appreciably above the experimental value.

As for the SciBooNE upper limit of the ratio of the $\pi^+$ 
coherent to total charged current cross-section, 
our prediction at $E_\nu$=1.1 GeV, which was 0.71$\cdot$10$^{-2}$, 
without $np-nh$ becomes 0.89$\cdot$10$^{-2}$, further above the experimental bound of 0.67$\cdot$10$^{-2}$.

%%%%%%%%%%%%%%%%%%%%%%%%%%%%%%%%%%%%%%%%%%%%%%%%%%%%%
\subsection{Quasi-elastic cross-section}
%%%%%%%%%%%%%%%%%%%%%%%%%%%%%%%%%%%%%%%%%%%%%%%%%%%%%%
\label{comp_qe}

\begin{figure}
\begin{center}
  \includegraphics[width=12cm,height=7cm]{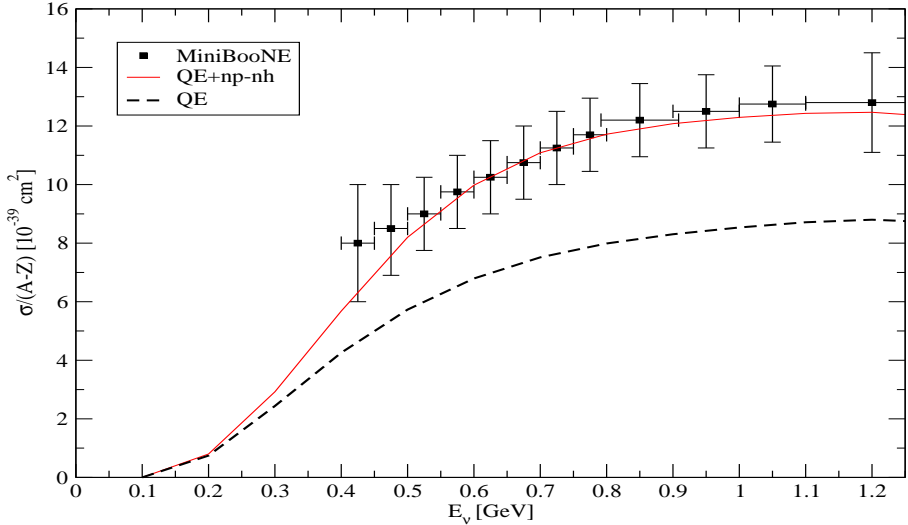}
\caption{``Quasi-elastic'' $\nu_\mu$-$^{12}$C cross-section per neutron 
as a function of neutrino energy. Dashed curve: pure quasi-elastic ($1p-1h$) cross-section; 
solid curve: with the inclusion of $np-nh$ component. The experimental MiniBooNE points 
are taken from \cite{Katori:2009du}.}
\label{fig_minib_qe}
\end{center}
\end{figure}

A new preliminary result on absolute cross-sections has been presented by the MiniBooNE collaboration 
\cite{Katori:2009du}. 
This group gives in particular the absolute value of the cross-section for
``quasi-elastic'' events, averaged over the neutrino flux and as a function of
neutrino energy. The comparison of these results with a prediction
based on the relativistic Fermi gas model using the standard value of the axial cut-off mass $M_A=1.03$ 
GeV$/c^2$
reveals a substantial discrepancy. In the same model a modification of the axial cut-off mass
from the standard value to the larger value $M_A=1.35$ GeV$/c^2$ is needed to account for data. 
A similar conclusion holds for the $Q^2$ distribution \cite{Gran:2006jn} 
\cite{:2007ru}.
The introduction of a realistic spectral function for the nucleon does not alter this conclusion 
\cite{Benhar:2009wi}.

As a possible interpretation we question here the real definition of 
quasi-elastic events.  As already discussed above, the nuclear medium is not a
gas of independent nucleons, correlated only by the Pauli principle, but there
are additional correlations.  The ejection of a single nucleon
(denoted as a genuine quasi-elastic event) is only one possibility, and one
must in addition consider events involving a correlated nucleon pair from
which the partner nucleon is also ejected. This leads to the excitation of
2 particle-2 hole ($2p-2h$) states which have been abundantly discussed throughout
this work. In the spin-isospin channel the correlations, mostly the tensor ones, 
add $2p-2h$ strength to the $1p-1h$ events \cite{Alberico:1983zg}. 
At present, in neutrino reactions, such events cannot be experimentally distinguished from
the genuine quasi-elastic events and must be considered simultaneously. 
Notice that the standard lower value of the axial mass, $M_A$=1.03 GeV$/c^2$,
results from deuterium bubble chamber experiments. 
In this case the effect of tensor correlation is also present but at a lower level since deuteron is 
a dilute system. Our
sum of the combined $^{12}$C quasi-elastic cross-section and the $2p-2h$ one is displayed
in Fig.\ref{fig_minib_qe}.  This prediction fits the experimental data excellently, 
better than expected in view of the  uncertainties of our $2p-2h$ cross-section. 
As for the flux averaged ``quasi-elastic'' cross-section per neutron 
the experimental value is $9.4 ~ 10^{-39}$ cm$^2$ (with a normalization error of $11 \%$). 
Our prediction for this quantity is $6.3 ~ 10^{-39}$ cm$^2$ 
without $2p-2h$ contribution and $9.0 ~ 10^{-39}$ cm$^2$ including it, a value more in touch 
with the experimental one.

\begin{figure}
\begin{center}
  \includegraphics[width=12cm,height=8cm]{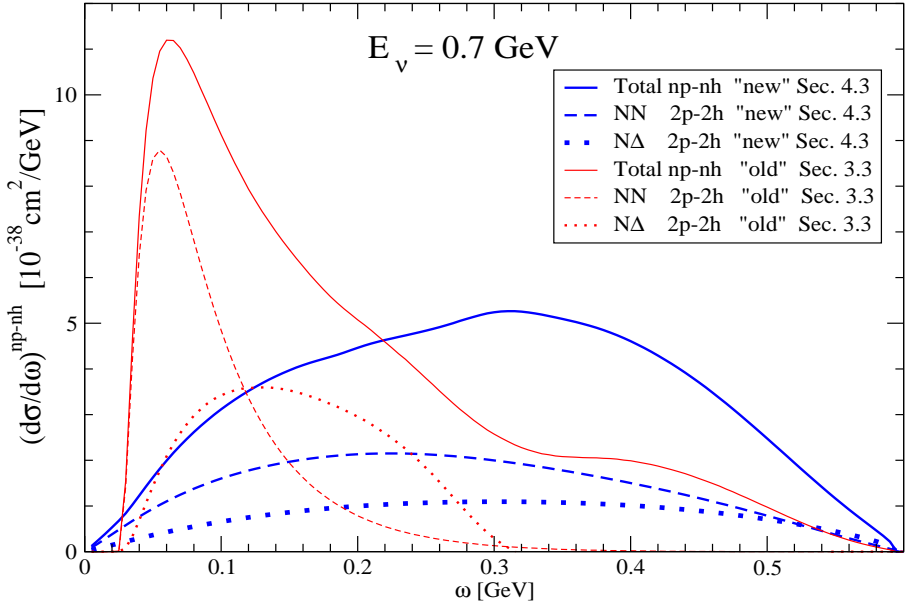}\\
\caption{Comparison between the CC $\nu_\mu$ -$^{12}$C $n p-n h~(n=2,3)$ 
differential cross-sections deduced from the two different parameterizations of the $2p-2h$ components. 
Thin lines: parametrization of Sec.\ref{subs_qe} and used throughout the whole article (denoted ``old''); 
bold lines: parametrization of Sec.\ref{comp_qe} (``new'').}
\label{fig_comp_npnh}
\end{center}
\end{figure}

In view of the importance of the issue 
we want to investigate if this large $2p-2h$ 
contribution is a genuine effect and not an artifact of the particular 
parametrization that we have used for the bare $2p-2h$ channel. For this, 
we introduce a different approach which exploits a microscopic evaluation 
by Alberico \textit{et al.} \cite{Alberico:1983zg} of the $2p-2h$ contribution to 
the transverse magnetic response of $(e,e')$ scattering. It does not have the shortcomings of our previous 
parameterizations which have no momentum dependence. 
In the previous case the maximum of the $2p-2h$ 
response $R^{NN}_{2p-2h}$ always lies at low energy, 
$\omega \simeq 50$ MeV, irrespective of the momentum, 
separating at large momentum from the
quasi-elastic peak which instead gets shifted at larger energies. 
A similar feature exists in the $N\Delta$ part. This is not realistic and 
below we sketch a possible way for improvements.
The aim is to extract the $2p-2h$ responses from the results of Alberico \textit{et al.} 
\cite{Alberico:1983zg}, 
although they are available 
for a limited set of momenta and energies and they concern iron instead of carbon. 
We have thus performed extrapolations both to cover 
all the kinematical region of neutrino reactions and to go to the $^{12}$C case. 
For the set of $R_{\sigma \tau (T)}(\omega,q)$ values that we could extract \cite{Alberico:1983zg} we have observed 
an approximate scaling behavior with respect to the variable $x=\frac{q^2-\omega^2}{2M_N \omega}$. 
A parametrization of the responses in terms of this variable allows 
the extrapolation needed to cover the full neutrino kinematical region and we
have now the new responses, $R^{NN}_{2p-2h}(\omega,q)$ and $R^{N\Delta}_{2p-2h}(\omega,q)$ in all the range. 
For the $\Delta \Delta$ part, which is not well covered in 
\cite{Alberico:1983zg} we have kept the previous parametrization, which already presents a proper 
$q$ dependence owing to the contribution of the in-medium $\Delta$ width \cite{Oset:1987re}.
Another remark is in order. 
The evaluation of Ref.\cite{Alberico:1983zg} of the $2p-2h$ channel
does not reproduce pion absorption in nuclei at threshold, as observed by the authors. 
It gives a too large value for the absorptive $p$-wave optical potential parameter \cite{Ericson:1988gk}, 
$\mathrm{Im} C_0 \simeq 0.18 m_\pi^{-6}$, instead of the best fit value  $\mathrm{Im} C_0 \simeq 0.11 m_\pi^{-6}$.
To be as consistent as possible with our previous parametrization, which comes from pion absorption, 
we have applied to our scaling function the reduction factor $\frac{0.11}{0.18}$. 
The nuclear mass dependence is taken care of with the introduction
of the Levinger factor,
$L$, which fixes the number of quasi-deuteron pairs in the nucleus defined as $L~ZN/A$.
We rescale 
the iron results by a factor $r$, ratio of the Levinger factors, for the two nuclei. It is $r=0.8$ according to the 
$A$ dependence of the Laget formula \cite{Laget:1981xg} or a similar value, $r\simeq0.75$ from \cite{Benhar:2003xr}.
Altogether the global reduction factor applied to the iron scaling function is $\simeq 0.5$.

\begin{figure}
\begin{center}
  \includegraphics[width=12cm,height=5cm]{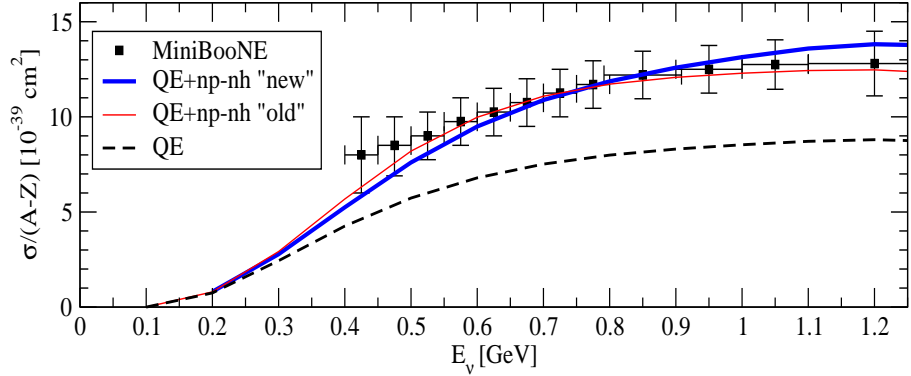}
\caption{Same as Fig.\ref{fig_minib_qe} including also 
our curve (QE + $np-nh$ ``new'') with the new parametrization for the $2p-2h$ component (bold line).}
\label{fig_comp_npnh_mini}
\end{center}
\end{figure}

Since in the previous case the RPA have little effect on the $2p-2h$ component, we introduce
directly the bare new $2p -2h$ quantities
 in the neutrino cross-section. 
The influence of the new modelization of the $2p-2h$ is displayed in Fig. \ref{fig_comp_npnh} where 
the bare partial and total $np-nh$ differential CC neutrino cross-sections at $E_\nu=0.7$ GeV 
are shown both for the previous parametrization and for the new one. The energy behaviors are quite different,
the $NN$ contribution is no longer localized at small energies but is spread over the whole energy range, 
a more realistic feature. A similar behavior occurs for $N \Delta$ part. 
However the integral over the energy, $\sigma_{np-nh}(E_\nu)$, is practically not modified. 
As a consequence adding 
this contributions to the quasi-elastic cross-section we reach a similar conclusion as before 
about the important role of the multi-nucleon channel, as illustrated in Fig. \ref{fig_comp_npnh_mini}.

It indicates that, in the nuclear medium, neutrino in this energy range 
do not interact only with individual nucleons 
but also with pairs of nucleons, mostly $n-p$ pairs correlated by the tensor interaction. 
The spin dependent part of the neutrino interaction  with such a pair is stronger 
than with the same two nucleons when isolated. 
This increase manifests itself trough the $2p-2h$ strength which adds to the $1p-1h$ part, an  effect 
simulated by an increase of the axial cut-off mass. 
Quantitatively a confirmation on the theoretical side of the exact magnitude through a 
detailed microscopic calculation of the bare $2p-2h$ 
response, which will then be inserted in our RPA formalism, would be helpful.
Also an experimental  identification of the final state would be of a great importance 
to clarify this point. In particular the charge of
 the ejected nucleons will be quite significant. 
Since tensor correlations involve $n-p$ pairs, the ejected pair is predominantly $p-p$ for
 charged current and $n-p$ for neutral current. This predominance has the same origin as for 
 p-wave $\pi^-$ absorption by nuclei where $n-n$ emission is favored over $n-p$ emission 
\cite{Shimizu:1980kb,Ericson:1966fm}.

%%%%%%%%%%%%%%%%%%%%%%%%%%%%%%%%%%%%%%%%%%%%%%%%%%%%%%%%%%%%%
\section{Summary and Conclusion}
%%%%%%%%%%%%%%%%%%%%%%%%%%%%%%%%%%%%%%%%%%%%%%%%%%%%%%%%% 

We have studied neutrino interactions with light nuclei 
which enter the targets of present or future experiments. 
Our theoretical tool is the theory of the nuclear response treated 
in the random phase approximation (RPA), a well established technique for the treatment of
electromagnetic or weak interactions with nuclei and which have been used 
also for strongly interacting probes. 
The crucial element of the RPA treatment is the $p-h$ interaction, in particular for the spin-isospin one, which 
has been taken from the accumulated knowledge on the spin-isospin responses.
The main merit of this approach is to allow unified description of various channels. 
It has some limitations which restrict the energy range of the neutrino to a region below $\simeq 1.2$ GeV. 
For instance the only nucleonic resonance incorporated in the description is the Delta resonance. 
Multi-pion production is also ignored, 
as well as most meson exchange effects. Moreover, although both the Delta propagator and the 
center-of-mass correction are the relativistic one, not all relativistic effects are included in a systematic way.

The final states considered  are the
 quasi-elastic, $2p-2h$, $3p-3h$ ones, and coherent or incoherent pion production.  
Some channels have  the problem that final state interactions are not incorporated.
 This is the case for incoherent pion emission where the produced pion can be absorbed on its way out
 of the nucleus leading to a multi-nucleon state. 
Incoherent pion production is therefore
 overestimated and the multi-nucleon channel accordingly underestimated.
This effect is visible in the scattering of physical pions on $^{12}$C in the region of the Delta peak. 
For a light nucleus such as $^{12}$C the effect is limited but it becomes more serious in heavier nuclei.
Our method should be supplemented by an evaluation of the final state interaction, 
for instance by a Monte Carlo method \cite{Leitner:2006ww},\cite{VicenteVacas:1994sm}.

 The coherent channel is particularly interesting
although it represents only a small fraction of the total pion emission.
 It does not suffer from the previous  limitations as final state interactions are automatically 
incorporated in the RPA treatment which is particularly suited for this channel. 
Moreover, it is the only channel which is dominated by the isospin spin-longitudinal 
response where collective effects are very pronounced
while they remain moderate in the other channels.
 The difference between the first order term with one bubble (with Delta excitation) and the full RPA chain is quite
large. In this context we have used as a test of our spin longitudinal response the experimental data on 
elastic pion scattering in the Delta region. It offers a direct test 
of the forward coherent neutrino cross-section to which 
it is linked through the Adler theorem. Except for low pion energies near threshold ($\omega \le 200$ MeV) 
where Adler's theorem fails, the elastic cross-section can be used to extract 
the forward neutrino coherent cross-section as in the method of Rein-Sehgal.

For the evolution of the partial cross-sections with mass number in order to reach the $^{40}$Ca region, 
our description indicates that, 
apart from the coherent pion production which evolves differently, 
the other partial cross-sections scale essentially as the nucleon number. 
Final state interactions obviously will destroy this scaling.

We have compared our predictions with the available experimental data.
Our ratio for the coherent pion production over the total neutrino cross-section is 
just compatible with the experimental upper limit. 
Another test concerns the ratio for charged currents $\pi^+$ production to the quasi-elastic cross-section. 
A delicate point in the experiments concerns the definition of a quasi-elastic process and 
its separation from  $np-nh$ which the experiment does not distinguish. 
In one set of data a correction is applied to obtain a genuine quasi-elastic cross-section and it is
corrected as well for final state interaction. 
In another set of results a generalized quasi-elastic is introduced, 
defined as events with only one lepton. 
In this case our $2p-2h$ and $3p-3h$ should be added to the quasi-elastic component. 
Both lead to  successful comparisons with the two sets of experimental data. 
Further data involve a ratio of neutral current $\pi^0$ production 
to the total neutrino cross-section for charged currents. 
Here again our evaluation agrees with data. 
It is quite encouraging that the comparison with present experimental data is essentially successful.

A distinct feature of our approach, and one of our significant results, 
is the large $2p-2h$ component. It affects all the measured ratios discussed in this work. 
At the present level of accuracy we have not found in these ratios 
any contradiction to its presence. 
It is also supported by preliminary data on the absolute neutrino quasi-elastic cross-section on carbon. 
We suggest that the proposed increase of the axial mass from the standard value to a larger one
to account for the quasi-elastic data, reflects the presence of a polarization cloud, mostly due to tensor interaction, 
which surrounds a nucleon in the nuclear medium. 
It translates into a final state with ejection of two nucleons, 
which in the present stage of the experiments is indistinguishable from the quasi elastic final state. 
Although the existence of such $2p-2h$ component is not in question,
for a fully quantitative evaluation
we plan to improve the description of the multi-nucleon
final states by a microscopic treatment. 
Future precision experiments, such as T2K, will be able to identify final states, 
namely $p-p$ pairs for charged current and $n-p$ pairs for neutral current,
and bring an experimental elucidation of this intriguing effect.

\vskip 0.5 true cm
\noindent {\bf \large{Acknowledgments}}

%\vskip 0.2 true cm
\noindent
We thank Luis Alvarez-Ruso, Dany Davesne, Torleif Ericson and Alfredo Molinari for stimulating discussions.

\appendix
%{\bf \Large{APPENDIX}}
\section{Inclusive neutrino-nucleus cross-section}
\label{appendix A}
The invariant amplitude for the lepton-nucleus cross-section, Eq. (\ref{m_eq_1}), results from the contraction 
between the leptonic $L$ and the hadronic $H$ tensors
\begin{eqnarray} \label{eq/amplitude}
|T|^2 &=& L_{00}W^{00} \, + \, L_{33}W^{33} \, + \, (L_{03}+L_{30})W^{03} \, + \nonumber \\ && (L_{11}+L_{22})W^{11} \, \pm \, 
(L_{12}-L_{21})W^{12} \,\,\, \begin{cases} + & \text{($\nu$)} \\ - & \text{($\bar{\nu}$)} \end{cases}.
\end{eqnarray}
The various $L$ are the component of the leptonic tensor
\begin{equation}
L_{\mu\nu} = 8 (k_\mu {k'}_\nu + k_\nu {k'}_\mu - g_{\mu\nu} k.k' \mp i \varepsilon_{\mu\nu\alpha\beta} k^\alpha {k'}^\beta)
\end{equation}
while the $W$ of the hadronic one
%\begin{equation}
%W^{\mu\nu} = \sum_{P,P'= N,\Delta} \sqrt{\frac{M'}{E'}}\sqrt{\frac{M'}{E'}} \, H^{\mu\nu}_{PP'}.
%\end{equation}
\begin{eqnarray}
W^{\mu\nu}& =& \sum_{P,P'= N,\Delta} \sqrt{\frac{M_P}{E_P^q}}\sqrt{\frac{M_{P'}}{E_{P'}^q}} \, H^{\mu\nu}_{PP'}=\nonumber\\
&=&\frac{M_N}{E_N^q} H^{\mu\nu}_{NN}+\sqrt{\frac{M_N}{E_N^q}}\sqrt{\frac{M_{\Delta}}{E_{\Delta}^q}}H^{\mu\nu}_{N\Delta}+
\frac{M_{\Delta}}{E_{\Delta}^q}H^{\mu\nu}_{\Delta\Delta},
\end{eqnarray}
where
 $E_{N}^{q} = (\gvec{q}^2+{M_N}^2)^{1/2}$ and $E_{\Delta}^{q} = (\gvec{q}^2+{M_{\Delta}}^2)^{1/2}$. 
This decomposition takes into account the different channels of particle-hole excitations.\\
The various leptonic tensor components are: 
\begin{eqnarray}
L_{00} &=& 8( k_0 {k'}_0 + kk'\cos\theta ), \nonumber \\
L_{33} &=& 8(2 k_3 {k'}_3 + k_0 {k'}_0 - kk'\cos\theta ), \nonumber \\
L_{03}+L_{30} &=& 16( k_0 {k'}_3 + k_3 {k'}_0 ), \nonumber \\
L_{11}+L_{22} &=& 16( 2 k_1^2+k_0 {k'}_0- k {k'}\cos\theta), \nonumber \\
L_{12}-L_{21} &=& - i~8 ( k_0 {k'}_3 - k_3 {k'}_0 ), \nonumber \\
\mathrm{with} && k_3 = \frac{k}{q}(k'\cos\theta - k), \nonumber \\
              && {k'}_3 = \frac{k'}{q}(k' - k\cos\theta), \nonumber \\
\mathrm{and} &&  k_1=k_2=\frac{kk'}{q}\frac{\sin \theta }{\sqrt{2}}.
\end{eqnarray}
For the hadronic tensor components we keep only the leading 
terms in the development of the hadronic current in $ p/M $, where $ p $ 
denotes the initial nucleon momentum. Marteau investigated the importance of the momentum terms and found them to be small.
The components are related to the various nuclear responses $R$ as follows 
\begin{eqnarray}
H^{00}_{PP'} &=& \alpha^0_P \alpha^{0}_{P'} R_{\tau} + \beta^0_P \beta^{0}_{P'} R_l\nonumber \\
H^{03}_{PP'} &=& \alpha^0_P \alpha^{3}_{P'} R_{\tau} + \beta^0_P \beta^{3}_{P'} R_l\nonumber \\
H^{33}_{PP'} &=& \alpha^3_P \alpha^{3}_{P'} R_{\tau} + \beta^3_P \beta^{3}_{P'} R_l\nonumber \\
H^{11}_{PP'} &=& \gamma^0_P \gamma^0_{P'} R_t+\delta^0_P\delta^0_{P'} R_t \nonumber \\
H^{22}_{PP'} &=& H^{11}_{PP'}, \nonumber \\
H^{12}_{PP'} &=& -i \gamma^0_P\delta^0_{P'} R_t - i \delta^0_P \gamma^0_{P'} R_t.
\end{eqnarray}
For sake of illustration we give the explicit expression of $H^{00}$:
\begin{eqnarray}
H^{00}&=&\sum_{P,P'= N,\Delta}H^{00}_{PP'}=\nonumber\\
&=&\alpha^0_N \alpha^0_{N} R_{\tau}^{NN}+
%\alpha^0_N \alpha^0_{\Delta} R_{\tau}^{N\Delta}+
%\alpha^0_{\Delta} \alpha^0_{\Delta} R_{\tau}^{\Delta\Delta}+
\beta^0_N \beta^0_{N} R_l^{NN}+2 \beta^0_N \beta^0_{\Delta} R_l^{N\Delta}+\beta^0_{\Delta} \beta^0_{\Delta} R_l^{\Delta\Delta}.\nonumber\\
\end{eqnarray}

The quantities $\alpha,\beta,\gamma$ and $\delta$ are expressed in terms of the usual form factors, namely
\begin{eqnarray}
\alpha_P^0 &=& N_{P}^{q} \left[ F_1 - F_2 \frac{\gvec{q}^2}{2M_P(E_{P}^{q}+M_P)} \right], \nonumber \\
\alpha_P^3 &=& N_{P}^{q} \left[ F_1 - F_2 \frac{\omega}{2M_N} \right] \frac{\nvec{q}}{E_{P}^{q}+M_P}, \nonumber \\
\beta_P^0 &=& N_{P}^{q} \left[ G_A^* - G_P \frac{\omega}{2M_N} \right] \frac{\nvec{q}}{E_{P}^{q}+M_P}, \nonumber \\
\beta_P^3 &=& N_{P}^{q} \left[ G_A - G_P \frac{\gvec{q}^2}{2M_N(E_{P}^{q}+M_P)} \right], \nonumber \\
\gamma_P^0 &=& N_{P}^{q} \left[ F_1 - F_2 \frac{\omega}{2M_N} + F_2 \frac{E_{P}^{q}+M_P}{2M_P} \right] \frac{\nvec{q}}{E_{P}^{q}+M_P}, 
\nonumber \\
\delta_P^0 &=& - N_{P}^{q} G_A.
\end{eqnarray}
 
We have introduced in the time component of the axial current 
a renormalization factor $G_A^*=G_A(1+\delta)$ to account meson exchange effects which are known to be 
important in this channel \cite{Kubodera:1978wr}. 
Even with the large value $\delta=0.5$ the effect of this renormalization is small. 
The most affected channel is the coherent, which is reduced by $\simeq10 \%$.

\subsection{Spin longitudinal contribution to the quasi elastic cross-section}
\label{appendix_long_QE}
We consider the limit of vanishing lepton mass. In this case the relevant leptonic tensor components reduce to
\begin{equation}
L_{00}=4 [(k+k')^2-q^2]=\frac{q^2}{\omega^2} L_{33}=-\frac{1}{2}\frac{q}{\omega}(L_{03}+L_{30}).
\end{equation}
The longitudinal contribution to the quantity $|T|^2$ involves
\begin{equation}
\label{long_T2}
{\beta^0_{N}}^2 L_{00}+{\beta^3_{N}}^2 L_{33}+\beta^0_{N}\beta^3_{N} (L_{00}+L_{33})=
{N_{N}^{q}}^2 G_A^2 L_{00} \left[ \frac{q^2}{(E_{N}^{q}+M_N)^2}+
\frac{\omega^2}{q^2}-2\frac{\omega}{q}\frac{\nvec{q}}{E_{N}^{q}+M_N} \right]. 
\end{equation}
Neglecting the struck nucleon momentum, the transferred energy $\omega$ in a quasi-elastic process is $\omega=E_N^q-M_N$, 
which implies the bracket on the r.h.s. of Eq.(\ref{long_T2}) to vanish.

\section{Particle-hole polarization propagators}
\subsection{Bare}
\label{appendix_bare}
In this Appendix we give the expressions of the bare particle-hole polarization propagators.
The nucleon-hole polarization propagator is the standard Lindhard function \cite{walecka}.\\
%\begin{eqnarray} 
%\Pi^0_{N-h} (\omega,\gvec{q}) &=& 4 \,\, \int \,\, \frac{d^3 k}{(2\pi )^3} \,\, \left[ \frac{\theta (\nvec{q+k} - k_F ) \theta (k_F - \nvec{k})}{\omega + \omega_{\gvec{k}} -
%\omega_{\gvec{q+k}} + i\eta }  \right. \nonumber \\ && \left. - 
%\frac{\theta(k_F - \nvec{q+k}) \theta(\nvec{k} - k_F )}{\omega + \omega_{\gvec{k}} - \omega_{\gvec{q+k}} - i\eta } \right]
%\end{eqnarray}
%with $\omega_{\gvec{k}}=\frac{\gvec{k}^2}{2M_N}$.
For the Delta-hole polarization propagator we use the relativistic expression
\begin{equation}
\Pi_{\Delta-h}(q) = \frac{32 \tilde{M}_{\Delta}}{9} \int \, \frac{d^3 k}{(2\pi)^3}
\theta(k_F - k) \left[ \frac{1}{s - {\tilde{M}_\Delta}^2 +i \tilde{M}_\Delta \Gamma_{\Delta}} 
- \frac{1}{u - {\tilde{M}_{\Delta}}^2} \right],
\end{equation}
where $s$ and $u$ are the Mandelstam variables. $\tilde{M}_{\Delta}=M_{\Delta}+40 (MeV) \frac{\rho}{\rho_0}$ 
is the mass of the $\Delta$ 
in the nuclear medium and
$\Gamma_{\Delta}$ is the in medium Delta width. The last two quantities are 
taken from \cite{Oset:1987re}.

For the $2p-2h$ polarization propagators we consider only the imaginary parts. 
Their expressions, which represent an extrapolation of threshold results of \cite{Shimizu:1980kb} are

\begin{eqnarray} \label{eq/pi2p2h}
Im(\Pi^0_{NN}) & = & 4\pi\rho^2 \frac{(2M_N+m_\pi)^2}{(2M_N+\omega)^2} \, C_1 \, \Phi_1(\omega) \, \left[ \frac{1}{\omega^2} \right] \nonumber \\
Im(\Pi^0_{N\Delta}) & = & -4\pi\rho^2 \frac{(2M_N+m_\pi)^2}{(2M_N+\omega)^2} \, C_2 \, \Phi_2(\omega) \, \mathrm{Re} \left[ \frac{1}{\omega(\omega -\tilde{M}_\Delta + M_N +i\frac{\Gamma_\Delta}{2})} \right. \nonumber \\
&& \left. +  \frac{1}{\omega(\omega +\tilde{M}_\Delta -M_N)} \right] \nonumber \\
Im(\Pi^0_{\Delta\Delta}) & = & -4\pi\rho^2 \frac{(2M_N+m_\pi)^2}{(2M_N+\omega)^2} \, C_3 \, \Phi_3(\omega) \, 
\left[ \frac{1}{(\omega +\tilde{M}_\Delta -M_N )^2} \right]. 
\end{eqnarray}
The $C_i$ constants are set to $C_1=0.045$, $C_2=0.08$, $C_3=0.06$, while the $\Phi_i(\omega)$ 
include phase space, pion and rho propagators.

\subsection{RPA}
\label{appendix_rpa}
Here we define the RPA expressions of the response functions for finite nuclei. 

First we introduce the projection of the bare propagators on 
the Legendre's polynomials $P_L$ through

\begin{eqnarray}
\label{pl}
\Pi^{0(L)}(\omega,q,q') &=& 2\pi \, \int \, du \, P_L(u) \Pi^{0}(\omega,\gvec{q},\gvec{q'}), \nonumber \\
\Pi^{0(L)}_{k_F(R)}(\omega,q,q') &=& 2\pi \, \int \, du \, P_L(u) \Pi^{0}_{k_F(R)}(\omega,\frac{\gvec{q}+\gvec{q'}}{2}),
\end{eqnarray}
where $q=\nvec{q}$, $q'=\nvec{q'}$, $u=\cos(\widehat{q},\widehat{q'})$.\\
Starting from Eqs.(\ref{eq:laktineh}) and (\ref{pl}), after some algebraic manipulations, one obtains

\begin{eqnarray}
{\Pi}^{0(L)}(\omega,q,q') & = & 4\pi \sum_{l_1,l_2} (2l_1 +1)
(2l_2 +1) \left( \begin{array}{ccc}
	           l_1 & l_2 & L \\
	           0 & 0 & 0 \\ 
                 \end{array} \right)^2 \nonumber \\
&& \times \int \, dR R^2 \, j_{l_1}(qR) \, j_{l_1}(q'R)
{\Pi}_{k_F(R)}^{0(l_2)}(\omega,q,q') 
\end{eqnarray}
with the usual three-$j$ symbol and $l$-order Bessel function $j_l(x)$.\\ 
This is the starting point for the calculations of isovector and spin-isospin response functions.

The free isovector (or charge) response function can be expressed through
\begin{equation}
R^{0NN}_{cc}(\omega,q) = -\frac{\mathcal{V}}{\pi} \sum_J \frac{2J+1}{4\pi} \mathrm{Im} \left[ \Pi^{0(J)}_{N-h}(\omega,q,q) \right].
\end{equation}
The RPA isovector response function
\begin{equation}
R_{cc}^{NN} (\omega,q) = -\frac{\mathcal{V}}{\pi} \mathrm{Im} \left[ \Pi_{cc_{NN}}(\omega,\gvec{q},\gvec{q}),
\right] = -\frac{\mathcal{V}}{\pi} \sum_J \frac{2J+1}{4\pi} \mathrm{Im} \left[ \Pi_{cc_{NN}}^{(J)}(\omega,q,q) \right],
\end{equation}
is obtained solving the following equation
\begin{equation}
\Pi_{cc_{NN}}^{(J)}(\omega,q,q') = \Pi^{0(J)}_{N-h}(\omega,q,q') + \int \frac{dk \, k^2}{(2\pi)^3} \, \Pi^{0(J)}_{N-h}(\omega,q,k) \, 
V^{NN}_c(k) \, \Pi_{cc_{NN}}^{(J)}(\omega,k,q').    \end{equation}
%We parametrize the particle-hole interaction in this channel $V^{NN}_c(k)$ through the Landau-Migdal parameter $f'=0.6$.

For the spin-isospin longitudinal and transverse responses we introduce the following quantities
\begin{eqnarray}
\Pi^{0(J)}_{ll_{PP'}}(\omega,q,q') = \sum_{L=J\pm1} a_{JL}^2 \, {\Pi}^{0(L)}_{PP'}(\omega,q,q'), 
\nonumber \\
\Pi^{0(J)}_{lt_{PP'}}(\omega,q,q') = \sum_{L=J\pm1} a_{JL}b_{JL} \, {\Pi}^{0(L)}_{PP'}(\omega,q,q'),
\nonumber \\ 
\Pi^{0(J)}_{tt_{PP'}}(\omega,q,q') = \sum_{L=J\pm1} b_{JL}^2 \, {\Pi}^{0(L)}_{PP'}(\omega,q,q'). 
\end{eqnarray}
where
\begin{eqnarray}
a_{JL} &=& \begin{cases} -\sqrt{\frac{J}{2J+1}} & \text{for L=J-1}, \\
                       \sqrt{\frac{J+1}{2J+1}} & \text{for L=J+1}. \end{cases} \nonumber \\ 
b_{JL} &=& \begin{cases} \sqrt{\frac{J+1}{2J+1}} & \text{for L=J-1}, \\
                         \sqrt{\frac{J}{2J+1}} & \text{for L=J+1}, \\
                         1 & \text{for L=J}. \end{cases} 
\end{eqnarray}
Note that, in general, for finite systems $\Pi^{0(J)}_{lt}\neq 0$.\\ 
The bare responses in a particular channel $k$ ($k=QE,2p-2h,...$) are given by
\begin{equation}
R^{0PP'}_{(k)\,xy}(\omega,q) = -\frac{\mathcal{V}}{\pi} \, \sum_J \frac{2J+1}{4\pi} \, \mathrm{Im} [\Pi^{0(J)}_{(k)xy_{PP'}}(\omega,q,q)],
\end{equation}
with $x,y = l,t$, referred to the longitudinal or transverse channel, and $PP'=N,\Delta$.

The second term of Eq.(\ref{SEPAR}) in the channel $k$, namely
\begin{equation} 
\mathrm{Im} \Pi_{(k)} = \left| 1 + \Pi \, V \right|^2 \, \mathrm{Im}\Pi^0_{(k)},
\end{equation}
with $\Pi$ the full polarization propagator, explicitly writes
\begin{eqnarray}
\Pi^{(J)}_{(k)xy_{PP'}}(\omega,q,q') &=& \Pi^{0(J)}_{(k)xy_{PP'}}(\omega,q,q') \nonumber \\ && + \int \, \frac{dp \, p^2}{(2\pi)^3} \sum_{QR \atop ww'} \Pi^{0(J)}_{(k)xw_{PQ}}(\omega,q,p) \, V^{QR}_{ww'}(p) \, \Pi^{(J)}_{w'y_{QP'}}(\omega,p,q') \nonumber \\ && + \int \, \frac{dp \,  p^2}{(2\pi)^3} \sum_{QR \atop ww'} (\Pi^{(J)}_{xw_{PQ}}(\omega,q,p) \, V^{QR}_{ww'}(p))^* \, \Pi^{0(J)}_{(k)w'y_{QP'}}(\omega,p,q') \nonumber \\ && + \int\int \, \frac{dp \,  p^2}{(2\pi)^3}\frac{dp' \, {p'}^2}{(2\pi)^3} \, \sum_{QQ'RR' \atop ww'zz'} \, (\Pi^{(J)}_{xw_{PR}}(\omega,q,p) \, V^{RQ}_{wz}(p))^* \nonumber \\ && \Pi^{0(J)}_{(k)zz'_{QQ'}}(\omega,p,p') \, V^{Q'R'}_{z'w'}(p') \, \Pi_{w'y_{R'P'}}(\omega,p',q'),        
\end{eqnarray}
where $x,y,w,w',z,z'=l$ or $t$ and 
$P,P',Q,Q',R,R' \, = \, N,\Delta$.\\
The solution of this equation leads to the corresponding response functions
\begin{equation}
R^{PP'}_{(k)\,xy}(\omega,q) = -\frac{\mathcal{V}}{\pi} \, \sum_J \frac{2J+1}{4\pi} \, \mathrm{Im} [\Pi^{(J)}_{(k)xy_{PP'}}(\omega,q,q)].
\end{equation}
In our calculations the maximum multipole 
number is set to $J=25$ which turns out to be sufficient 
to reach the convergence.

The first term of Eq.(\ref{SEPAR}), which represents coherent processes, explicitly writes

\begin{eqnarray}
\Pi^{(J)}_{(co.)_{xyPP'}}(\omega,q,q') &=& \int \, \frac{dp\,p^2}{(2\pi)^3} \, 
(\Pi^{(J)}_{xlPQ}(\omega,q,p))^* \mathrm{Im} \left( V^{QQ'}_\pi(p) \right) \Pi^{(J)0}_{lyQ'P'}(\omega,p,q') \nonumber \\
&=& -i \, \frac{q_\pi^2}{16\pi^2} \left( \frac{f^2}{m_\pi^2} \, (\Pi^{(J)}_{xlPN}(\omega,q,q_\pi))^* \, \Pi^{(J)}_{lyNP'}(\omega,q_\pi,q') \right. \nonumber \\ && + \, \frac{ff^*}{m_\pi^2} \, (\Pi^{(J)}_{xlPN}(\omega,q,q_\pi))^* \, \Pi^{(J)}_{ly\Delta P'}(\omega,q_\pi,q') \, \nonumber \\ &&+ \, \frac{f^*f}{m_\pi^2} \, (\Pi^{(J)}_{xlP\Delta}(\omega,q,q_\pi))^* \, \Pi^{(J)}_{lyNP'}(\omega,q_\pi,q') \nonumber \\ && \left. + \, \frac{f^{*2}}{m_\pi^2} \, (\Pi^{(J)}_{xlP\Delta}(\omega,q,q_\pi))^* \, \Pi^{(J)}_{ly\Delta P'}(\omega,q_\pi,q') \right),
\end{eqnarray}
where
\begin{equation}
\mathrm{Im} \left( V_\pi \right)= \mathrm{Im} 
\left(C_\pi \frac{{\gvec{q}}^2}{\omega^2 -{\gvec{q}}^2 -m_\pi^2 + i\eta}\right)=
 -i \, C_\pi \, \pi \, \gvec{q}^2 \delta(\gvec{q}^2 - q_\pi^2) =-i \, C_\pi \, \pi \, \frac{q_\pi}{2} \, \delta(\nvec{q}-q_\pi), 
\end{equation}
with $C_\pi$ the generic Nucleon- or Delta-pion coupling constant and $q_\pi = \sqrt{\omega^2 - m_\pi^2}$.

\end{document}